\documentclass{article}

\usepackage{arxiv}

\usepackage[utf8]{inputenc} 
\usepackage[T1]{fontenc}    
\usepackage{url}            
\usepackage{amsthm,amsmath,amsfonts,amssymb,bm,float,dirtytalk,subcaption,cite,graphicx,xcolor}

\newcommand{\bbeta}{\bm{\beta}}
\newcommand{\bsigma}{\bm{\sigma}}
\newcommand{\Tr}{{\rm Tr}}
\newcommand{\rme}{{\rm e}}
\newcommand{\rmd}{{\rm d}}
\theoremstyle{plain}
\newtheorem{thm}{Theorem}[section]
\newtheorem{result}[thm]{Result}

\title{Correction of overfitting bias via leave one out theory}

\author{
 Emanuele Massa \\
  DCN, Donders Institute\\
  Radboud University\\
  Nijmegen, The Netherlands\\
  \texttt{emanuele.massa@donders.ru.nl} \\
   \And
    Marianne A. Jonker\\
    Section Biostatistics, Department for Health Evidence\\
    Radboud UMC\\
    Nijmegen, The Netherlands\\
  \texttt{marianne.jonker@radboudumc.nl} \\
  \And
 Anthony C.C. Coolen \\
 DCN, Donders Institute\\
  Radboud University, \\
  Saddle Point Science Europe\\
  Nijmegen, The Netherlands
  \texttt{a.coolen@science.ru.nl} \\
}

\begin{document}
\maketitle
\begin{abstract}
Regression analysis based on many covariates is becoming increasingly common. However, when the number of covariates $p$ is of the same order as the number of observations $n$, maximum likelihood regression becomes unreliable due to overfitting. This typically leads to systematic estimation biases and increased estimator variances. It is crucial for inference and prediction to quantify these effects correctly. Several methods have been proposed in literature to overcome overfitting bias or adjust estimates. The vast majority of these focus on the regression parameters. 
But failure to estimate correctly also the nuisance parameters may lead to significant errors in confidence statements and outcome prediction. 

In this paper we present a jacknife method for deriving a compact set of non-linear equations which describe the statistical properties of the ML estimator in the regime where $p=O(n)$ and under the hypothesis of normally distributed covariates. These equations enable one to compute the overfitting bias of maximum likelihood (ML) estimators in parametric regression models as functions of $\zeta = p/n$. We then use these equations to compute shrinkage factors in order to remove the overfitting bias of maximum likelihood (ML) estimators. 
This new derivation offers various benefits over the replica approach in terms of increased transparency and reduced assumptions.  To illustrate the theory we performed simulation studies for multiple regression models. In all cases we find excellent agreement between theory and simulations.
\end{abstract}


\section{Introduction}
\label{intro}
When the number of parameters included in a statistical model is large compared to the number of observations, the noise (or residual model variance) is wrongly attributed to the deterministic part  of the underlying model. This is called overfitting \cite{babyak, harrel1, johnstone}. When a model \say{overfits} the data, the resulting estimators are generally affected by bias and are subject to large sample-to-sample fluctuations. This is a problem in inference, as one generally seeks (asymptotically) unbiased estimators with small variance. It is also a problem for prediction: an overfitting model may appear to perform well for metrics based on the training data, i.e.\ data used to fit the model, but will fail when used to predict outcomes for new data.
In modern applications, regression models frequently include a large number of regression parameter;  this is why 
correction for overfitting in regression models has become an important research topic \cite{coolen2017,GLM,vandegeer, johnstone,hastie,steyerberg}.

We consider the following setting: we are given $n$ independent observations $(T_1,\mathbf{X}_1), \dots, (T_n,\mathbf{X}_n)$ of response $T\in \mathbb{R}$ and covariates $\mathbf{X}\in \mathbb{R}^p$. The responses are assumed to have been generated by the model
\begin{equation}
     T|\mathbf{X} \sim f(.|\mathbf{X}'\bbeta_0,\bsigma_0),
     \label{eq:form}
\end{equation}
where we indicate with $'$ the transpose operation and $(\bbeta_0 \in \mathbb{R}^p$, $\bsigma_0 \in \mathbb{R}^s)$ are the true parameter values. When $p<n$, these parameters are usually estimated with the maximum likelihood (ML) estimators
\begin{equation}
    \hat{\bbeta}_n,\hat{\bsigma}_n := \underset{\bbeta,\bsigma}{\arg\max} \big\{\ell_n(\bbeta,\bsigma)\big\},
\end{equation}
where the log-likelihood function $\ell_n(\bbeta,\bsigma)$ is defined as 
\begin{eqnarray}
     \ell_n(\bbeta,\bsigma) := \frac{1}{n} \sum_{i=1}^n \log f (T_i|\mathbf{X}'_i\bbeta,\bsigma).
\end{eqnarray}
In the classical regime $p\ll n$, and under regularity conditions, the ML estimator $(\hat{\bbeta}_n, \hat{\bsigma}_n)$ concentrates around the true values $(\bbeta_0, \bsigma_0)$. 
When $p=O(n)$ this is generally no longer true and the ML estimators are typically affected by bias and subject to large sample-to-sample fluctuations (i.e. large variance). 
These two features are the fingerprints of overfitting \cite{babyak}.
In literature one finds various rules of thumb on the maximum number of covariates that can be included in a regression analysis to prevent overfitting \cite{peduzzi2, harrell2}. But in fields like medicine, the number of observations is often small and models may be complex, which makes such restrictions on the number of covariates highly undesirable.
Another common strategy is to fit a penalized model, i.e. add a suitable penalization function to the log-likelihood \cite{harrell2, steyerberg}. Finally, a simple post-hoc procedure to combat overfitting effects is to shrink the regression coefficients after estimation \cite{Copas, steyerberg}. The penalization parameter or the shrinking factor can be estimated via bootstrapping, by selecting the value which minimizes a certain error measure \cite{steyerberg}. But this must be done with care in the setting $p=O(n)$ as it has been shown that bootstrapping is not always reliable \cite{purdom}.
Furthermore, measures of overfitting bias are practically non-trivial to compute as we do not have access to \say{good} estimators of the true parameters.  

A possible way out is to model the behaviour of the ML estimator as a function of the true parameters, which can be done effectively only under hypotheses on the data generating distribution. In this paper we follow this line of thought and study the average behaviour of the ML estimator under the hypothesis that  
\begin{equation}
\label{gauss_law}
    \mathbf{X} \sim \mathcal{N}(\bm{0},\bm{\Sigma}_0), \quad  \bm{\Sigma}_0 \in\mathbb{R}^{p\times p}
\end{equation}
with positive definite covariance matrix $\bm{\Sigma}_0\succ 0$ and such that 
\begin{equation}
    \theta_0 := \|\bm{\Sigma}_0^{1/2}\bbeta_0\| = O(1).
\end{equation}
These assumptions and the fact that there exists an interesting link between statistical physics and optimization \cite{mezard, kirkpatrick, virasoro} enable tools from statistical physics to be applied to statistical inference \cite{loureiro, barbier_hd, donoho}. Several papers have exploited this \say{bridge} in the last decade, thoroughly characterizing the behaviour of the estimator $\hat{\bbeta}_n$ for the regression coefficients. However, to the best of our knowledge, less attention has been paid to understanding the behaviour of $\hat{\bsigma}_n$, i.e. the estimator of the nuisance parameters which do not enter in the model via a linear combination of the covariates. These parameters must also be estimated to construct confidence intervals for the regression parameters or prediction intervals for the outcome variable and might be meaningful in the statistical analysis of the data (see Subsection \ref{subsection:weibull} for an example in medical statistics).   

In this paper we show that the ML estimators of regression and nuisance parameters can be simultaneously corrected for overfitting bias, in order to obtain unbiased estimators even in the proportional asymptotic regime $p = O(n)<n$. In doing so, we also provide a novel and more intuitive derivation of the results of \cite{coolen2017,GLM,sheikh,massa},  which used statistical physics methods to estimate the statistical relation between $\hat{\bbeta}_n,\hat{\bsigma}_n$ and $\bbeta_0,\bsigma_0$ for generalized linear models (GLM). 
Our derivation is based on the-leave-one-observation-out approach of \cite{el_karoui2013}. This procedure can be traced back to the jacknife \cite{jacknife}, and resembles the cavity method from statistical physics \cite{Mezard_89,massa}. It offers the advantage of being based on explicit approximations, rather than on the algebraic identity used in the replica method \cite{coolen2017,GLM}. Furthermore it allows us to understand the limitations of previous and present results \cite{coolen2017,GLM,sheikh,massa,el_karoui2013} that arise when $\bbeta_0$ is not \say{diffuse}, i.e.\ when each component of $\bbeta_0$ is no longer scaling as $1/\sqrt{p}$. 

This paper is organized as follows. In Section \ref{section:theory} we present our  theory for estimating the asymptotic bias for models of the family (\ref{eq:form}). 
For those models that via a suitable transformation can be written in a form that is linear in the regression coefficients, we show that although $\hat{\bbeta}_n$  is always asymptotically unbiased, $\hat{\bsigma}_n$ is still biased. For non-linear models, both $\hat{\bbeta}_n$ and $\hat{\bsigma}_n$ are always biased. 
In Section \ref{section:applications} we illustrate  the theory via  application to the Log-logistic AFT model for time to event analysis, the Weibull Proportional Hazards Model for time to event data \cite{kalbfleisch}, and the Logit regression model for binary outcome data. We confirm via numerical simulations how our theory can be used to correct ML estimates of both regression and nuisance parameters reliably for overfitting bias. We end the paper with a discussion of our results and future directions of investigation. Most mathematical derivations are delegated to appendices.

\section{Maximum Likelihood estimator in the overfitting regime}
\label{section:theory}

In Subsection \ref{sub:representation} we introduce a stochastic representation for $\hat{\bbeta}_n$, which basically tells us that, under our hypotheses, the distribution of $\hat{\bbeta}_n$ is completely characterized by two scalar random variables $K_n$ and $V_n$, with a precise geometrical meaning.
We also introduce the approximations $\xi_i$ and  $\bar{\bsigma}_n$ for  the linear predictors $\mathbf{X}_i'\hat{\bbeta}_n$ and $\hat{\bsigma}_n$ respectively. Assuming that these approximations are sufficiently accurate, we show in Subsection \ref{sub:RS_eqs} that we can derive a set of equations for the limiting values of $K_n, V_n$ and $\hat{\bsigma}_n$, which hold when the sample size $n$ and the number of covariates $p$ tend to infinity with fixed ratio $\zeta=p/n$. These limiting values have direct interpretations and allow to quantify the bias and more generally the behaviour of the ML estimator in the relevant asymptotic limit.

\subsection{Expressions for $\hat{\bbeta}_n$ and $\hat{\bsigma}_n$}
\label{sub:representation}

When $\mathbf{X}$ follows a Multivariate Normal distribution (with a non-singular covariance matrix), it has been shown  \cite{sur_covariance,massa}  that by the rotational invariance of the multivariate standard normal distribution and the particular conditional dependence of the response $T|\mathbf{X}= T|\mathbf{X}'\bbeta_0$, we have the representation 
\begin{equation}
\label{representation_beta}
     \hat{\bbeta}_n\big(\bbeta_0,\bsigma_0,\{(T_i,\mathbf{X}_i)_{i=1}^n\}\big) \overset{d}{=} K_n \bbeta_0 + V_n \bm{\Sigma}_0^{-1/2}\mathbf{U}, 
\end{equation}
which is valid at arbitrary $n$ and $p$, where
\begin{eqnarray}
\label{K_n}
    K_n &=&  (\bbeta_0'\bm{\Sigma}_0\hat{\bbeta}_n)/\|\bm{\Sigma}_0^{1/2}\bbeta_0\|^2 \\
\label{V_n}
    V^2_n &=& \hat{\bbeta}_n'\bm{\Sigma}_0\hat{\bbeta}_n -K_n^2\|\bm{\Sigma}_0^{1/2}\bbeta_0\|^2.
\end{eqnarray}
and $\mathbf{U}$ is uniformly distributed on the unit sphere $\mathcal{S}_{p-2}$ of $\mathbb{R}^{p-1}$ that is orthogonal to $\bm{\Sigma}_0^{1/2}\bbeta_0$ (for details 
see Appendix \ref{appendix:representation}).  
Representation (\ref{representation_beta}) is very useful: it implies that the statistical behaviour of $\hat{\bbeta}_n$ is completely characterized by the two  random variables $K_n$ and $V_n$. Virtually any statistical property of interest of the estimator $\hat{\bbeta}_n$ can be expressed in terms of (the moments of) $K_n, V_n$ and $\bm{\Sigma}_0$. 
For instance, the squared bias reads
\begin{equation}
\label{bias_beta}
    \|\mathbb{E}[\hat{\bbeta}_n]-\bbeta_0\|^2 = (\mathbb{E}[K_n]-1)^2\|\bbeta_0\|^2.
\end{equation}

The ML estimators $(\hat{\bbeta}_n,\hat{\bsigma}_n)$ are defined as 
\begin{equation}
    (\hat{\bbeta}_n,\hat{\bsigma}_n) = \underset{\bbeta,\bsigma}{\arg\max} \ \ell_n\big(\bbeta,\bsigma\big).
\end{equation}
Hence  $\hat{\bbeta}_n$ and $\hat{\bsigma}_n$ are the solution of the score equations:
\begin{eqnarray}
\label{beta_estimator}
     \nabla_{\bbeta}\ell_n (\bbeta,\bsigma) &=& \frac{1}{n}\sum_{i=1}^n \mathbf{X}_i \dot{u}(\mathbf{X}_i'\hat{\bbeta}_n,T_i,\bsigma) = 0\\
\label{nuis_estimator}
    \nabla_{\bsigma}\ell_n (\bbeta,\bsigma) &=& \frac{1}{n}\sum_{i=1}^n \mathbf{g}(\mathbf{X}_i'\hat{\bbeta}_n,T_i,\bsigma) = 0, 
\end{eqnarray}
where $u(x,y,\mathbf{z}):=\log f(y|x,\mathbf{z})$, $\dot{u}(x,y,\mathbf{z}) = \partial_{x}u(x,y,\mathbf{z})$ and where $ \mathbf{g}(x,y,\mathbf{z}) := \nabla_{\mathbf{z}} u (x,y,\mathbf{z})$.
We henceforth indicate with $\hat{\bbeta}_n(\bsigma)$ the solution of (\ref{beta_estimator}) at fixed $\bsigma$.
Note that (\ref{nuis_estimator}) depends on $\hat{\bbeta}_n(\bsigma)$ only through the linear predictors $\mathbf{X}_i'\hat{\bbeta}_n(\bsigma)$. In Appendix \ref{appendix:subsection_lp} we show that $\mathbf{X}_i'\hat{\bbeta}_n(\bsigma)$ can be approximated as 
\begin{equation}
\label{lp_approx}
    \mathbf{X}_i'\hat{\bbeta}_n(\bsigma) \simeq \xi_i:= {\rm prox}_{-\tau_n u(.,T_i,\bsigma)}\big(\mathbf{X}_i'\hat{\bbeta}_{(i)}(\bsigma)\big).
\end{equation}
Here $ {\rm prox}_{g}$ denotes the proximal mapping of a convex function $g:\mathbb{R}\rightarrow \mathbb{R}$, 
\begin{equation}
     {\rm prox}_{g}(x) := \underset{y}{\arg\min}\Big\{\frac{1}{2}(y-x)^2 + g(y)\Big\},
\end{equation}
$\hat{\bbeta}_{(i)}(\bsigma)$ is the leave $i$-th out version of $\hat{\bbeta}_n(\bsigma)$ (the solution of (\ref{beta_estimator}) upon neglecting the $i$-th observation at fixed $\bsigma$) and $\tau_n= \tau_n(\zeta,\bsigma)$ is the solution of the following equation (see Appendix \ref{appendix:subsection_tau}):
\begin{equation}
\label{eq_tau1}
    1-\zeta  =  \frac{1}{n}\sum_{i=1}^n \frac{1}{1-\tau_n \ddot{u}(\xi_i,T_i,\bsigma) }, 
\end{equation}
where $\ddot{u}(x,y,z) = \partial^2_x u(x,y,z)$.
Upon using (\ref{lp_approx}) in (\ref{nuis_estimator}), we obtain 
\begin{equation}
\label{nuis_approx}
    \nabla_{\bsigma}\ell_n (\hat{\bbeta}_n(\bsigma),\bsigma)  \simeq \frac{1}{n} \sum_{i=1}^n \mathbf{g}(\xi_i,T_i,\bsigma).
\end{equation}
It then follows that one can obtain $\tau_n$ and an approximation $\bar{\bsigma}_n$ for $\hat{\bsigma}_n$ and $\mathbf{X}_i'\hat{\bbeta}_n$, by solving (\ref{eq_tau1}, \ref{nuis_approx}) simultaneously for $\bsigma$ and $\tau$.

\subsection{Asymptotic self-consistency equations}
\label{sub:RS_eqs}
Note that $K_n$ and $V_n$ in (\ref{K_n}, \ref{V_n}) are sums of $p$ random variables, the components of $\hat{\bbeta}_n$. If these are independent, the central limit theorem guarantees that $K_n$ and $V_n$ asymptotically concentrate around their means. If they are correlated, more advanced versions of the theory of concentration of measure must be used to establish the concentration rate of $K_n$ and $V_n$. 
In what follows we shall assume that as $n, p \rightarrow \infty$, any correlations present are sufficiently weak for the variance of $K_n$ and $V_n$ to vanish  asymptotically, i.e.\ that at fixed $\zeta$ and $\bsigma$:
\begin{eqnarray}
    K_n &=& k(\zeta,\bsigma)+ o_P(1)\label{k_concentration}\\
    V_n &=& v(\zeta,\bsigma)+o_P(1),\label{v_concentration} \
\end{eqnarray}
We will verify our assumption a posteriori via numerical simulations.
Defining $\tilde{\bbeta}_0 = \bm{\Sigma}_0^{1/2}\bbeta_0$, we have 
\begin{equation}
    \mathbf{X}_i'\hat{\bbeta}_{(i)} \overset{d}{=}
    \bm{\mathcal{X}}_i'\frac{\Tilde{\bbeta}_0\Tilde{\bbeta}_0'}{\|\Tilde{\bbeta}_0\|^2}\bm{\Sigma}_0^{1/2}\hat{\bbeta}_{(i)} + \bm{\mathcal{X}}_i'\Big(\bm{I} - \frac{\Tilde{\bbeta}_0\Tilde{\bbeta}_0'}{\|\Tilde{\bbeta}_0\|^2}\Big)\bm{\Sigma}_0^{1/2}\hat{\bbeta}_{(i)} , 
\end{equation}
where we used 
\begin{equation}
    \mathbf{X}_i \overset{d}{=} \bm{\Sigma}_0^{1/2}\bm{\mathcal{X}}_i, \quad ~~\bm{\mathcal{X}}_i\sim \mathcal{N}(\bm{0},\bm{I}_{p\times p}).
\end{equation}
Since for
\begin{eqnarray}
    Z_{0,i} & :=& \frac{\bm{\mathcal{X}}_i'\Tilde{\bbeta}_0}{\|\Tilde{\bbeta}_0\|} \sim \mathcal{N}(0,1)\\
    \mathbf{Q}_i &:=&\Big(\bm{I} - \frac{\Tilde{\bbeta}_0\Tilde{\bbeta}_0'}{\|\Tilde{\bbeta}_0\|^2}\Big) \bm{\mathcal{X}}_i\sim\mathcal{N}(\bm{0},\bm{I}_{p-1})
\end{eqnarray}
with $\mathbf{Q}_i  \perp Z_{0,i}$ we also have that 
\begin{equation}
\label{lp}
    \mathbf{X}_i'\hat{\bbeta}_{(i)} \overset{d}{=} \frac{\Tilde{\bbeta}_0'\bm{\Sigma}_0^{1/2}\hat{\bbeta}_{(i)}}{\|\Tilde{\bbeta}_0\|} Z_{0,i} + \mathbf{Q}_i'\Big(\bm{I} - \frac{\Tilde{\bbeta}_0\Tilde{\bbeta}_0'}{\|\Tilde{\bbeta}_0\|^2}\Big)\bm{\Sigma}_0^{1/2}\hat{\bbeta}_{(i)}
    \overset{d}{=} K_{(i)}\theta_0 Z_{0,i} + V_{(i)}Q_i 
\end{equation}
with $\theta_0 := \|\Tilde{\bbeta}_0\|$,  $Q_i \sim \mathcal{N}(0,1)$ and where $K_{(i)},V_{(i)}$ are the leave the $i$-th observation out versions of (\ref{K_n}, \ref{V_n}).
Hence from (\ref{k_concentration}, \ref{v_concentration}) we infer that at fixed $\bsigma$ and $\zeta$
\begin{equation}
\label{lp_approx_2}
    \mathbf{X}_i'\hat{\bbeta}_{(i)} \overset{d}{=}  k\theta_0 Z_{0,i}  + vQ_i. 
\end{equation}
Using the approximation (\ref{lp_approx_2}) together with (\ref{lp_approx}) leads to
\begin{equation}
\label{score_beta}
    \nabla_{\bbeta} \ell_n(\bsigma,\bbeta)= \frac{1}{n}\sum_{i=1}^n \mathbf{X}_i \dot{u}(\mathbf{X}_i'\hat{\bbeta},T_i,\bsigma)\simeq \frac{1}{n}\sum_{i=1}^n \mathbf{X}_i \dot{u}(\xi_i,T_i,\bsigma) =0,
\end{equation}
where 
\begin{equation}
    \xi_i := {\rm prox}_{-\tau u(.,T_i,\bsigma)}(k\theta_0 Z_{0,i}  + vQ_i).
\end{equation}
Projecting (\ref{score_beta}) onto $\bbeta_0$ and using (\ref{lp_approx_2}), we obtain 
\begin{equation}
\label{score_k}
    \frac{1}{n}\sum_{i=1}^n \frac{\bm{\mathcal{X}}_i'\Tilde{\bbeta}_0}{\|\Tilde{\bbeta}_0\|^2}\big(\xi_i  - k\theta_0 Z_{0,i}  - vQ_i\big)=0.
\end{equation}
Similarly, projecting (\ref{score_beta}) onto $\hat{\bbeta}_n$, we get another equation
\begin{equation}
\label{score_v}
     \frac{1}{n}\sum_{i=1}^n \xi_i \ \dot{u}\big(\xi_i,T_i,\bsigma\big) = 0.
\end{equation}
The left-hand sides of (\ref{score_k}, \ref{score_v}) are both sums of i.i.d. random variables, which will converge to their mean in probability, provided both have a finite variance, 
\begin{eqnarray}
    \label{eq_k}
     &&\frac{1}{n}\sum_{i=1}^n \frac{\bm{\mathcal{X}}_i'\Tilde{\bbeta}_0}{\|\Tilde{\bbeta}_0\|^2}\Big(\xi_i  - (k\theta_0 Z_{0,i}  + vQ_i)\Big)\xrightarrow[]{P}\mathbb{E}\Big[ Z_0\Big(\xi - k\theta_0 Z_{0}  -vQ\Big)\Big] \\
     \label{eq_v}
     &&\frac{1}{n}\sum_{i=1}^n  \xi_i \ \dot{u}\Big(\xi_i,T_i,\bsigma\Big)~\xrightarrow[]{P}~\mathbb{E}\Big[ \xi \dot{u}\Big(\xi,T,\bsigma)\Big)\Big] 
\end{eqnarray}
where
\begin{equation}
    \xi := {\rm prox}_{-\tau u(.,T,\bsigma)}(k\theta_0 Z_{0}  + vQ).
\end{equation}
The right-hand side of (\ref{nuis_approx}) will similarly converge at fixed $\bsigma$ to its expectation
\begin{equation}
\label{eq_sigma}
     \frac{1}{n} \sum_{i=1}^n \mathbf{g}\big(\xi_{i},T_i,\bsigma \big) ~\xrightarrow[]{P}~ \mathbb{E}\Big[ \mathbf{g}\big({\rm prox}_{-\tau u(.,T,\bsigma)}(k(\bsigma)\theta_0 Z_{0}  + v(\bsigma))Q),T,\bsigma \big)\Big].
\end{equation}
The same reasoning implies that 
\begin{equation}
\label{eq_tau}
   \frac{1}{n} \sum_{i=1}^n\frac{1}{1-\tau \ddot{u}(\mathbf{X}_i'\hat{\bbeta}_n,T_i,\bsigma) } ~ \xrightarrow[]{P}~ \mathbb{E}\Big[\frac{1}{1-\tau \ddot{u}(\xi,T,\bsigma) } \Big].
\end{equation}
After few algebraic simplifications that can be found in Appendix \ref{appendix:subsection_self_consistent_eqs}, we see that $K_n, V_n, \tau_n$ and $\bar{\bsigma}_n$ of (\ref{nuis_approx}) will converge for $n\to\infty$ to the limiting (deterministic) values $k_{\star},v_{\star},\tau_{\star}$ and $\bsigma_{\star}$, that satisfy the following coupled nonlinear equations:
\begin{result}
[Replica Symmetric equations]
\label{result:mainresult}
    \begin{eqnarray}
        \label{rs1}
        \zeta v^2 &=& \tau^2\mathbb{E}_{T,Z_0,Q}\Big[\Big(\dot{u}(\xi,T,\bsigma)\Big)^2\Big]\\
        \label{rs2}
        1-\zeta &=& \mathbb{E}_{T,Z_0,Q}\Big[\frac{1}{1-\tau \ddot{u}(\xi,T,\bsigma) }\Big]\\
        \label{rs3}
        k \theta_0 &=& \mathbb{E}_{T,Z_0,Q}\Big[ Z_0\xi\Big] \\
        \label{rs4}
         {\bm{0}} &=& \mathbb{E}_{T,Z_0,Q}\Big[ \mathbf{g}(\xi,T,\bm{\sigma})\Big]
    \end{eqnarray}
    Here $ u(x,y,\mathbf{z}):=\log f(y|x,\mathbf{z})$, $\mathbf{g}(x,y,\mathbf{z}) := \nabla_{\mathbf{z}} u (x,y,\mathbf{z})$,  $\theta_0 := \|\bm{\Sigma}_0^{1/2}\bbeta_0\| = O_n(1)$,  $Z_0,Q \sim \mathcal{N}(0,1), \ Z_0\!\perp\! Q,\ T|Z_0 \sim p(.|\theta_0Z_0,\bm{\sigma}_0)$, and we use the short-hand
    \begin{equation}
    \label{RS_eqs}
        \xi = {\rm prox}_{-\tau u(.,T,\bsigma)}(k \theta_0Z_0+vQ) \ .
    \end{equation}
\end{result} 
\noindent
These equations for the quantities $(k_\star,v_\star,\tau_\star,\bsigma_\star)$ that characterize the asymptotic quantitative features of overfitting GLM regression models are identical to the results of the replica analysis in \cite{GLM}, but are now obtained in a completely different way (without any assumptions on the distribution of $\bbeta_0$). They are only dependent on: (i) the modulus of the effective association $\theta_0 := \|\bm{\Sigma}_0^{1/2}\bbeta_0\|$, (ii) the ratio $\zeta = p/n$, and (iii) the {true}  nuisance parameters $\bsigma_0$. Upon inversion, they enable us to create {\em unbiased} estimators for the regression parameters and the nuisance parameters, expressed in terms of the {\em biased} ML-inferred values.

\subsection{The special case of linear models}

If the model under consideration depends linearly on the covariates, i.e.\ 
\begin{equation}
    T_i = \mathbf{X}_i'\bbeta + \epsilon_i, \qquad \epsilon \sim f_{\epsilon}(.|\bsigma)
\end{equation}
then, as was already shown in \cite{el_karoui2013}, one will have $K_n=1$ in (\ref{representation_beta}). So  the regression coefficients will be unbiased. 
To see this within the present formalism one may simply introduce the short-hand 
$\hat{\Delta}_n  = \hat{\bbeta}_n-\bbeta_0$ to write the ML estimate
\begin{eqnarray}
    \hat{\bbeta}_n &=& \underset{\bbeta}{\arg\max}\Big\{\underset{\bsigma}{\max}\Big\{\frac{1}{n} \sum_{i=1}^n \log f_\epsilon(T_i-\mathbf{X}_i'\bbeta,\bsigma)\Big\}\Big\}
    \end{eqnarray}
in the alternative form 
 \begin{eqnarray}
\label{Linear_delta}
    \hat{\bm{\Delta}}_n &:=& \underset{\bm{\Delta} }{\arg\max}\Big\{\underset{\bsigma}{\max}\Big\{\frac{1}{n} \sum_{i=1}^n \log f_\epsilon\big(\epsilon_i-\mathbf{X}_i'\bm{\Delta} ,\bsigma\big)\Big\}\Big\} \
\end{eqnarray}
with $\epsilon_i = T_i-\mathbf{X}_i'\bbeta_0$ independent of $\mathbf{X}_i$.
The latter is a regression problem with true association vector $\bm{\Delta}_0= \bm{0}$, so according to  (\ref{representation_beta}), we get 
\begin{equation}
\label{repr_linear}
    \hat{\bbeta}_n = \bbeta_0 + V_n\bm{\Sigma}_0^{-1/2}\mathbf{U} 
\end{equation}
where 
\begin{equation}
    V^2_n := \|\bm{\Sigma}_0^{1/2}(\hat{\bbeta}_n-\bbeta_0)\|^2\\
\end{equation}
and $\mathbf{U}$ is uniformly distributed on the unit sphere $\mathcal{S}_{p-1}$ of $\mathbb{R}^{p}$. 
In (\ref{K_n}) we now have $K_n = 1$, so $\hat{\bbeta}_n$ is unbiased for all $n$, because $\mathbb{E}[\mathbf{U}]=\bm{0}$. 
For the transformed problem (\ref{Linear_delta}) where $\bm{\Delta}_0= \bm{0}$, we can simply obtain the asymptotic value $v^\star =\lim_{n\to\infty} \mathbb{E}[V_n]$ from the RS equations (\ref{rs1}--\ref{rs4}) upon setting $\theta_0 = \|\bm{\Sigma}_0^{1/2}\bbeta_0\| = 0$.
The same must then be true for the untransformed problem.
We now find that (\ref{RS_eqs}) is independent of $Z_0$,
  \begin{equation}
    \label{RS_eqs_lin}
        \xi = {\rm prox}_{-\tau u(.,T,\bsigma)}\big(vQ\big)
    \end{equation}
    and hence (\ref{rs3}) is satisfied trivially. 
  
\vspace*{3mm}

Linear models for which  $\epsilon$ obeys a distribution in a location-scale family (see Appendix \ref{appendix:location_scale}) are of particular interest. They are defined by the property
\begin{equation}
    \epsilon \overset{d}{=}  \phi + \sigma Z,  \quad Z \sim f_Z
\end{equation}
for some fixed distribution $f_Z$, and with $\sigma\geq 0$. 
We show in Appendix \ref{appendix:location_scale} that for such models the RS equations (\ref{rs1}-\ref{rs4}) can always be re-written as
\begin{eqnarray}
       \label{lin:rs1}
        \zeta v^2/\sigma^2 &=&\mathbb{E}_{Z,Q}\Big[ \big(\Tilde{\xi}-v/\sigma Q-(\phi-\phi_0)/\sigma + \sigma_0/\sigma Z\big)^2\Big] \\
        \label{lin:rs2}
       1-\zeta &=& \mathbb{E}_{Z,Q}\Big[\frac{1}{1-\tilde{\tau}\ddot{\tilde{u}}(-\tilde{\xi})}\Big]\\
        \label{lin:rs3}
        \frac{\phi-\phi_0}{\sigma} &=& \mathbb{E}_{Z,Q}\Big[\Tilde{\xi}\Big] + (\sigma_0/\sigma)\mathbb{E}\Big[Z\Big] \\
        \label{lin:rs4}
        \frac{\sigma}{\sigma_0} &=&-\mathbb{E}_{Z,Q}\Big[Z\dot{\tilde{u}}(-\Tilde{\xi})\Big]
\end{eqnarray}
where $Q\sim\mathcal{N}(0,1)$ and 
 \begin{equation}
\label{lin:rs5}
    \Tilde{\xi} =- {\rm prox}_{-\Tilde{\tau} \Tilde{u}}\big((\phi_0-\phi)/\sigma + \sigma_0/\sigma Z-v/\sigma Q\big)
\end{equation}
with $\Tilde{\tau}:=\tau/\sigma^2$, $\Tilde{u} := \log f_Z$. 
Besides a welcome simplification compared to the more general case, this important result tells us that the four quantities  $(\phi_{\star}\!-\!\phi_0)/\sigma_{\star}$, $\sigma_{\star}/\sigma_0$,  $v_{\star}/\sigma_{\star}$ and $\Tilde{\tau}_{\star}$ are universal, i.e.\ they do not depend on $\phi_0$ or $\sigma_0$,  but only on $\zeta$. Hence, to solve the RS equations we do not need to know the full generative model underlying the data, but only the distribution $f_Z$.

\section{Application to selected regression models}
\label{section:applications}

In this section we test the accuracy of the following asymptotic approximation of (\ref{representation_beta}), which is obtained upon replacing $(K_n,V_n)\to \lim_{n\to\infty} (K_n,V_n)=(k_\star,v_\star)$,   
\begin{equation}
\label{asymp_representation}
     \hat{\bbeta}_n\approx k_\star \bbeta_0 + v_\star \bm{\Sigma}_0^{-1/2}\mathbf{U}, 
\end{equation}
against simulated data for regression models of interest in reliability analysis and time-to-event analysis.
In particular, we show for three different regression models how the theory can be used very effectively to compute bias correction factors for both the regression and the nuisance parameters. 
In Subsection \ref{subsection:linear} we study a linear model in the location scale family with application in survival analysis, namely the accelerated failure time log-logistic model.
Subsections \ref{subsection:weibull} and \ref{subsection:logit} deal with more complicated nonlinear models (i.e.\ models where $T$ depends nonlinearly on the covariates), the Weibull and the logit regression model. Here both $\hat{\bbeta}_n$ and $\hat{\bsigma}_n$ are asymptotically biased and the RS equations do generally depend on $\theta_0=\|\Tilde{\bbeta}_0\| = \|\bm{\Sigma}_0^{1/2}\bbeta_0\|$. The Weibull model represents a very special case of a non-linear model where the RS equations do not depend on $\|\Tilde{\bbeta}_0\|$. The logit model presents a rich phenomenology and the behaviour of the estimator $\hat{\bbeta}_n$, as well as its existence have been extensively studied in recent years \cite{HD_logit}. 
We extend these previous results by introducing an intercept term in the model and by providing a numerical routine that computes the desired correction factors by taking as input only measurable quantities.
In this case we need to estimate the true values $\|\Tilde{\bbeta}_0\|$ and $\phi_0$ that appear in the RS equations. We present a strategy for the estimation of the intercept, which has been so far not considered in previous studies on the argument. Furthermore we study by numerical simulations the distribution of the estimators obtained by these numerical procedures. 

Covariate correlations can be incorporated trivially, 
so we will for simplicity  consider  $\mathbf{X}\sim \mathcal{N}(\bm{0},{\bf I})$, i.e.\ $\bf{\Sigma}_0={\bf I}$. Note that global properties like the log-likelihood density, $k_{\star}$ and $v_{\star}$, depend on $\bbeta_0$ only through $\theta_0 = \|\tilde{\bbeta}_0\| = \|\bm{\Sigma}_0^{1/2}\bbeta_0\|$. In contrast, the distribution of the components of $\hat{\bbeta}_n$ does depend on that of $\bbeta_0$, not only on $\theta_0$. 
In all the following examples, we will choose  $\bbeta_0 = \theta_0\mathbf{e}_1$.  Here we expect from our theory that all components $\mathbf{e}_k'\hat{\bbeta}_n$ for $k\geq2$ will asymptotically have the same zero-average distribution, but that  $\mathbf{e}_1'\hat{\bbeta}_n$ will have a non zero average and a larger variance.

\subsection{Log-Logistic AFT model}
\label{subsection:linear}

\begin{figure}[H]
\begin{subfigure}{.49\textwidth}
  \includegraphics[width=\linewidth]{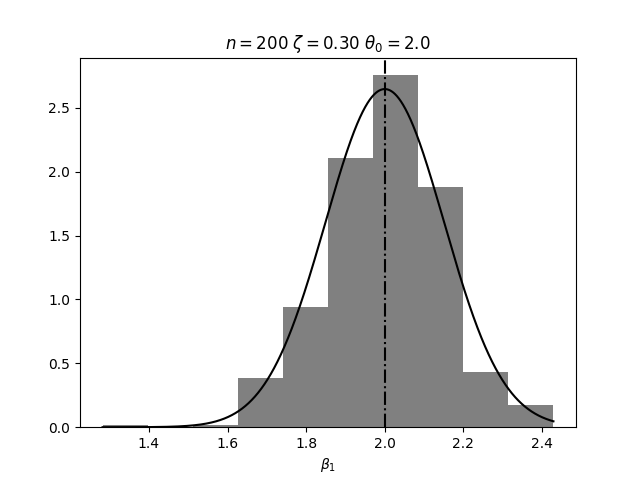}
  \caption{}
  \label{fig:1a}
\end{subfigure}
\hfill
\begin{subfigure}{.49\textwidth}
  \includegraphics[width=\linewidth]{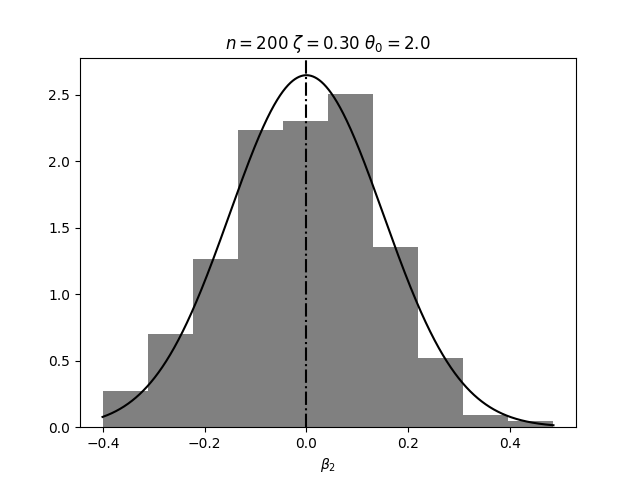}
  \caption{}
  \label{fig:1b}
\end{subfigure}
\\
\begin{subfigure}{.5\textwidth}
  \includegraphics[width=\linewidth]{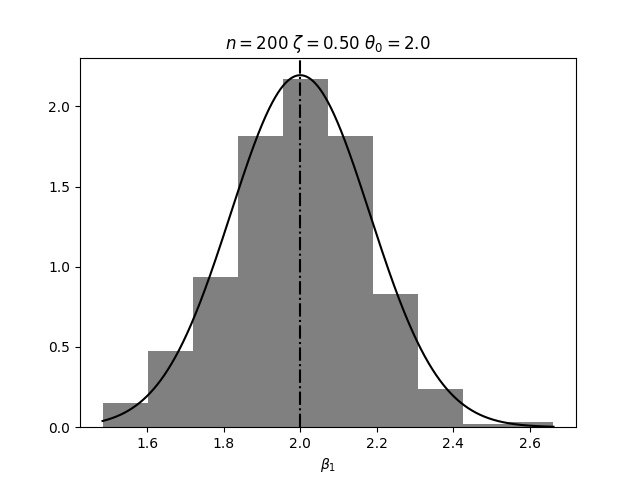}
  \caption{}
  \label{fig:1c}
\end{subfigure}%
\hfill
\begin{subfigure}{.5\textwidth}
  \includegraphics[width=\linewidth]{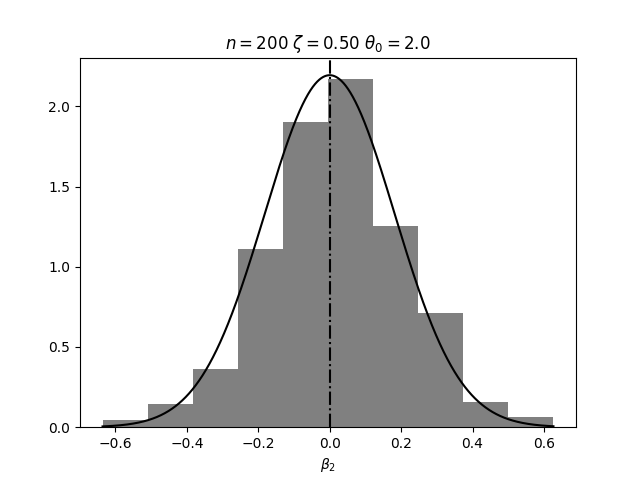}
  \caption{}
  \label{fig:1d}
\end{subfigure}
\caption{Simulated data for the Log-logistic model (\ref{log_linear}). The histograms of $m=500$ realizations of first (left) and second  (right) components of $\hat{\bbeta}_n$, each based on $n=200$ i.i.d. samples, for two different values of $\zeta = 0.3$  (\ref{fig:1a},\ref{fig:1b}) and $\zeta =0.5$  (\ref{fig:1c},\ref{fig:1d}). Dashed line: true value of the parameter, $\bbeta_0 =\theta_0 \mathbf{e}_1$. 
Solid line: the density corresponding to the representation (\ref{repr_linear}), which is asymptotically $\hat{\bbeta}_n \approx \mathcal{N}(\bbeta_0,v_{\star}/p)$ . 
}
\label{fig:beta_log_logistic}
\end{figure}

The Log-logistic regression model is one of the most commonly adopted Accelerated Failure Time (AFT) models \cite{kalbfleisch} and is defined by 
\begin{equation}
    T|\mathbf{X} \sim  f(T|\mathbf{X}) = \frac{\rho\rme^{\phi}\big(T \rme^{\mathbf{X}'\bm{\beta}+\phi}\big)^{\rho-1}}{\big(1+(T \rme^{\mathbf{X}'\bm{\beta}+\phi })^{\rho}\big)^2}  .
    \label{AFT}
\end{equation} 
AFT models are often used if the proportional hazards assumption appears not to hold  \cite{kalbfleisch, harrell2}, and offer advantages over the Cox model in the interpretation of  regression parameters  \cite{harrell2}. They can be mapped to a linear model, since on logarithmic scale the response $Y= -\log T$ can be written as
\begin{equation}
\label{log_linear}
    Y = -\log(T) = \mathbf{X}'\bbeta + \phi + \sigma Z,  \qquad Z \sim f_Z(z) =  \frac{\exp(-z)}{\big(1+\exp(-z)\big)^2} \ .
\end{equation}
where  $\rho = 1/\sigma$.
Moreover, since the transformed AFT model is in the location-scale family, 
\begin{equation}
    \epsilon := \phi+\sigma Z, \quad \epsilon \sim f_\epsilon(\varepsilon)=\frac{1}{\sigma}f_{Z}\big(\frac{\varepsilon-\phi}{\sigma}\big),
\end{equation}
we can use the simplified RS equations (\ref{lin:rs1}, \ref{lin:rs2}, \ref{lin:rs3}).
For convenience we define 
\begin{equation}
    \chi(x)  := - {\rm prox}_{-\Tilde{\tau} \Tilde{u}}(x)/2= \frac{1}{2} x  - \frac{1}{2}\Tilde{\tau} \tanh\big(\chi(x)\big).
\end{equation} 
and note that both
 $\chi(.)$ and $- {\rm prox}_{-\Tilde{\tau} \Tilde{u}}(.) $ are anti-symmetric functions of $x$. 
Since $\mathbb{E}[Z] = 0$ we obtain
\begin{equation}
    (\phi-\phi_0)/\sigma = -\mathbb{E}_{Z,Q}[\Tilde{\xi}].
\end{equation}
Upon noting that 
\begin{eqnarray}
    \mathbb{E}_{Z,Q}[\Tilde{\xi}] &=& \mathbb{E}_{Z,Q}[- {\rm prox}_{-\Tilde{\tau} \Tilde{u}}\big(\sigma_0/\sigma Z - v/\sigma Q\big)] \nonumber \\
    &=&  \mathbb{E}_{Z,Q}[- {\rm prox}_{-\Tilde{\tau} \Tilde{u}}\big(-\sigma_0/\sigma Z + v/\sigma Q\big)]=\mathbb{E}_{Z,Q}[-\Tilde{\xi}],
\end{eqnarray}
which implies $ \mathbb{E}[\Tilde{\xi}]=0$, we then find that $\phi_{\star}=\phi_0$ is a solution of our equations, irrespective of the values of $v_{\star}$ and $\sigma_{\star}$.
After some simple algebraic manipulations (see Appendix \ref{appendix:log_log}) we obtain the following simplified expressions for our remaining RS equations that we must solve to obtain $(v_\star,\sigma_\star,\tilde{\tau}_{\star})$:
\begin{eqnarray}
    v^2/\sigma^2 \zeta &=& \Tilde{\tau}^2 \mathbb{E}\Big[\tanh^2(\chi_{\star})\Big] \\
     \Tilde{\tau} &=& \zeta\bigg(\mathbb{E}\Bigg[ \frac{1}{2\cosh^2(\chi_{\star}) +  \Tilde{\tau}}\Bigg]\bigg)^{-1}\\
    \sigma/\sigma_0&=& -\mathbb{E}\Big[Z\tanh(\chi_{\star})\Big]
\end{eqnarray}
where we denoted 
\begin{equation}
    \chi_{\star} = \chi\big(\sigma_0/\sigma Z - v/\sigma Q\big).
\end{equation}
In Figure \ref{fig:beta_log_logistic} we compare the approximate asymptotic representation (\ref{asymp_representation}) with the actual histogram of the estimator $\hat{\bbeta}_n$ for $m= 500$ samples, each consisting of $n = 200$ observations. The estimator of both components of $\hat{\bbeta}_n$ is unbiased and its variance equals $v_{\star}/\sqrt{p}$. This is in agreement with the representation (\ref{asymp_representation}), which reduces to 
\begin{equation}
    \hat{\bbeta}_n \approx \bbeta_0 + v_\star \bm{\Sigma}_0^{-1/2} \mathbf{U}
\end{equation}
with $\mathbf{U}$ uniformly distributed on the unit sphere $\mathcal{S}_{p-1}$ of $\mathbb{R}^{p}$. This is a consequence of (\ref{repr_linear}), due to the fact that the model is linear.

\begin{figure}[H]
\begin{subfigure}{.47\textwidth}
  \includegraphics[width=\linewidth]{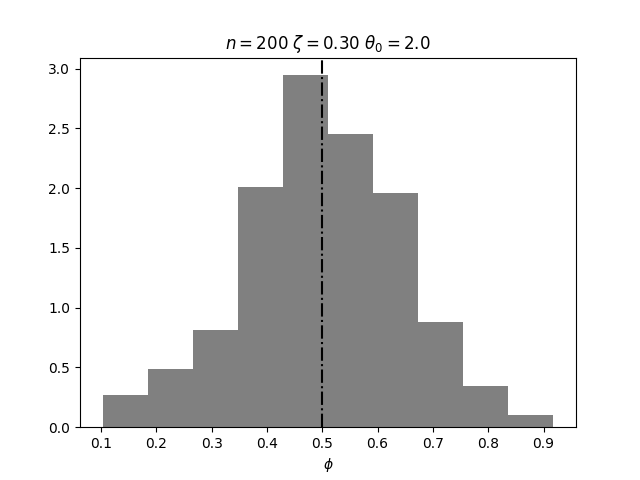}
  \caption{}
  \label{fig:2a}
\end{subfigure}%
\hfill
\begin{subfigure}{.47\textwidth}
  \includegraphics[width=\linewidth]{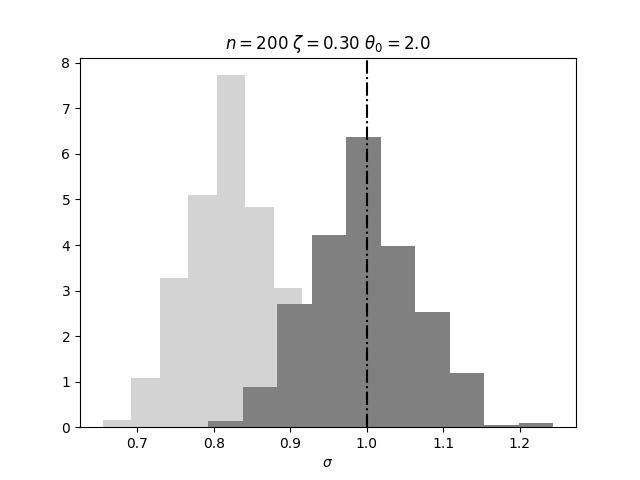}
  \caption{}
  \label{fig:2b}
\end{subfigure}
\medskip
\begin{subfigure}{.47\textwidth}
  \includegraphics[width=\linewidth]{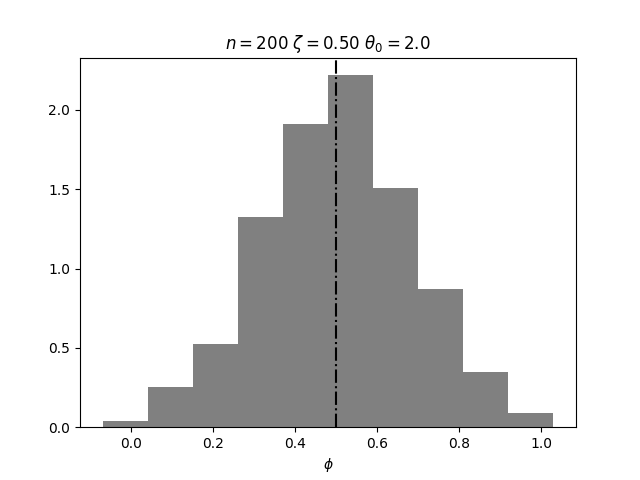}
  \caption{}
  \label{fig:2c}
\end{subfigure}%
\hfill
\begin{subfigure}{.47\textwidth}
  \includegraphics[width=\linewidth]{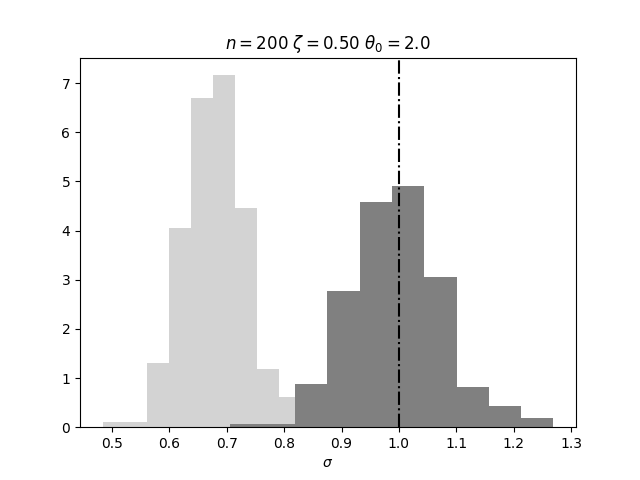}
  \caption{}
  \label{fig:2d}
\end{subfigure}
\caption{Simulated data for the Log-logistic model (\ref{log_linear}). The histograms of $m=500$ realizations of $\hat{\phi}_n$ (left) and $\hat{\sigma}_n$ (right, light grey) together with its corrected version $\tilde{\sigma}_n$ (right dark grey), each based on $n=200$ i.i.d.\ samples, for two different values of $\zeta = 0.3$  (\ref{fig:2a},\ref{fig:2b}) and $\zeta =0.5$  (\ref{fig:2c},\ref{fig:2d}). Dashed line: true value of the parameters $\phi_0=0.5$, $\sigma_0 = 1.0$ respectively. 
In this case we see that $\hat{\phi}_n$ is unbiased. On the other hand $\hat{\sigma}_n$ is clearly biased, while $\tilde{\sigma}_n$ is peaked around the true value.}
\label{fig:nuis_log_logistic}
\end{figure}

The estimators for the nuisance parameters are asymptotically biased, even if $\hat{\bbeta}_n$ is not. We can however use the fact that $\hat{\sigma}_n$ concentrates around the deterministic values $\sigma_{\star}$, hence the \say{corrected} estimator
\begin{equation}
    \tilde{\sigma}_n = \frac{\sigma_0}{\sigma_{\star}}\hat{\sigma}_n
\end{equation}
where $\sigma_{\star}/\sigma_0$ is the solution of the RS equations, will be approximately unbiased for $\sigma_0$. 
In Figure \ref{fig:nuis_log_logistic} we compare the histogram of the estimator $\hat{\sigma}_n$, with its corrected counterparts $\tilde{\sigma}_n$. The estimator $\hat{\phi}_n$ is unbiased, in agreement with the solution of the RS equations.
We cannot, as yet, predict the variance of the histograms of $\hat{\phi}_n$ and $\hat{\sigma}_n$, but we can already appreciate the fact that the estimates $\tilde{\sigma}_n$ are now centered around the true value $\sigma_0$.   
The correction factor $\sigma_{\star}/\sigma_0$ turns out to be always positive and less than $1$. This agrees with our intuition that when a model \say{overfits} it mistakenly explains the noise as part of the model. In turn this implies that the estimated noise width,  i.e.\ $\hat{\sigma}_n$, is progressively underestimated as the model gets more complex, i.e.\ $\zeta$ grows towards 1.
As a consequence the variance of the corrected estimator $\tilde{\sigma}_n$ is larger than the one of $\hat{\sigma}_n$.

\subsection{Weibull model}
\label{subsection:weibull}

The Weibull model is one of the most widely known models for skewed data \cite{kalbfleisch}, or time to event analysis and reads
\begin{equation}
\label{weib_model}
    T|\mathbf{X} \sim f(T|\mathbf{X}) = \rho_0 T^{\rho_0-1}\rme^{\mathbf{X}'\bbeta_0+\phi_0} \exp\Big\{-T^{\rho_0} \rme^{\mathbf{X}'\bbeta_0+\phi_0}\Big\}.
\end{equation}
Note that this implies 
\begin{equation}
    Z := T^{\rho_0} \rme^{\mathbf{X}'\bbeta_0+\phi_0} \sim {\rm Exp}(1) 
\end{equation}
or equivalenlty 
\begin{equation}
    T \overset{d}{=} Z^{1/\rho_0}\rme^{-\frac{\phi_0 + \mathbf{X}'\bbeta_0}{\rho_0}}.
\end{equation}
The log-likelihood associated with this model is 
\begin{eqnarray}
\label{ll_weibull}
    u(\mathbf{X}'\bbeta,T,\phi,\rho) &=& \log \rho + \rho \log T + \mathbf{X}'\bbeta+\phi -  \rme^{\rho\log T + \mathbf{X}'\bbeta+\phi}\\
    &\overset{d}{=}& \log \rho + \rho/\rho_0 \log Z +\phi- \rho/\rho_0\phi_0 + \mathbf{X}'\big(\bbeta -\rho/\rho_0\bbeta_0\big)  \nonumber \\
    &~~~~& - \rme^{\rho/\rho_0 \log Z +\phi- \rho/\rho_0\phi_0 + \mathbf{X}'\big(\bbeta -\rho/\rho_0\bbeta_0\big)}.
\end{eqnarray}
Hence the proximal mapping of the minus log-likelihood, 
when viewed as a function of the linear predictor $\mathbf{X}'\bbeta$, satisfies
\begin{eqnarray}
  z(x) - x  &=& \tau - \tau  \rme^{\rho/\rho_0 \log Z +\phi- \rho/\rho_0\phi_0 + x -\rho/\rho_0 \mathbf{X}'\bbeta_0}\rme^{z(x)}\\
  z(.) &=&  {\rm prox }_{-\tau u(.,T,\phi, \rho)}(.)
\end{eqnarray}
which can be formally solved by means of the Lambert W-function \cite{lambert_function}, which satisfies $W_0(x)\exp \big(W_0(x)\big) = x$, as 
\begin{equation}
    {\rm prox }_{-\tau u(.,T,\phi, \rho)}(x) = x + \tau -  W_0\Big(\tau \rme^{\tau + \rho/\rho_0 \log Z +\phi- \rho/\rho_0\phi_0 + x -\rho/\rho_0 \mathbf{X}'\bbeta_0}\Big).
\end{equation}
After some algebraic manipulations (see Appendix \ref{appendix:weibull}) we obtain the following set of RS equations
\begin{eqnarray}
    v^2 \zeta &=& \mathbb{E}_{Z,\Delta,Z_0,Q}\Bigg[ \bigg(\tau-  W_0\Big(\tau \rme^{\tau + \rho/\rho_0 \log Z +\phi- \rho/\rho_0\phi_0 + vQ} \Big)\bigg)^2\Bigg]\\
    \zeta &=& \mathbb{E}_{Z,\Delta,Z_0,Q}\Bigg[\frac{ W_0\Big(\tau \rme^{\tau + \rho/\rho_0 \log Z +\phi- \rho/\rho_0\phi_0 + vQ} \Big)}{1 +  W_0\Big(\tau \rme^{\tau + \rho/\rho_0 \log Z +\phi- \rho/\rho_0\phi_0 + vQ} \Big)} \Bigg]\\
    k &=& \rho/\rho_0\\
    \tau &=& \mathbb{E}_{Z,\Delta,Z_0,Q}\Big[W_0\Big(\tau \rme^{\tau + \rho/\rho_0 \log Z +\phi- \rho/\rho_0\phi_0 + vQ} \Big)\Big]\\
    \rho_0/\rho &=& \gamma_E +  \frac{1}{\tau}\mathbb{E}_{Z,\Delta,Z_0,Q}\Big[\log Z \ W_0\Big(\tau \rme^{\tau + \rho/\rho_0 \log Z +\phi- \rho/\rho_0\phi_0 + vQ} \Big)\Big].
\end{eqnarray}
Note that this system suggests, at a first glance, to treat $\phi_{\star}- \rho_{\star}/\rho_0\phi_0$ as the control parameter and this was indeed the way in which these equations were first solved. However it is easy to invert numerically the equation  
\begin{equation}
    \zeta = \mathbb{E}_{Z,\Delta,Z_0,Q}\Bigg[\frac{ W_0\big(\tau \rme^{\tau + \rho/\rho_0 \log Z +\phi- \rho/\rho_0\phi_0 + vQ} \big)}{1 +  W_0\big(\tau \rme^{\tau + \rho/\rho_0 \log Z +\phi- \rho/\rho_0\phi_0 + vQ} \big)} \Bigg]
\end{equation}
for instance by means of Newton's method. Let us use the definition $\Tilde{\phi} = \phi- \rho/\rho_0\phi_0$, at fixed $v,\rho/\rho_0,\tau$ we have
\begin{equation}
    \frac{\partial \zeta}{\partial \phi} = \mathbb{E}_{Z,\Delta,Z_0,Q}\Bigg[\frac{ W_0\big(\tau \rme^{\tau + \rho/\rho_0 \log Z +\phi- \rho/\rho_0\phi_0 + vQ} \big)}{\Big(1 +  W_0\big(\tau \rme^{\tau + \rho/\rho_0 \log Z +\phi- \rho/\rho_0\phi_0 + vQ} \big)\Big)^3} \Bigg]
\end{equation}
and the Newton iterate can be computed relatively easily
\begin{equation}
    \Tilde{\phi}^{t} = \Tilde{\phi}^{t-1} - \frac{\mathbb{E}_{Z,\Delta,Z_0,Q}\Bigg[\frac{ W_0\big(\tau \rme^{\tau + \rho/\rho_0 \log Z +\Tilde{\phi}^{t-1}  + vQ} \big)}{\Big(1 +  W_0\big(\tau \rme^{\tau + \rho/\rho_0 \log Z +\Tilde{\phi}^{t-1} + vQ} \big)\Big)^3} \Bigg]}{ \mathbb{E}_{Z,\Delta,Z_0,Q}\Bigg[\frac{ W_0\big(\tau \rme^{\tau + \rho/\rho_0 \log Z +\Tilde{\phi}^{t-1} + vQ} \Big)}{1 +  W_0\big(\tau \rme^{\tau + \rho/\rho_0 \log Z +\Tilde{\phi}^{t-1} + vQ} \big)} \Bigg]-\zeta} \ .
\end{equation}
We then obtain a new system of RS equations that can be solved by fixed point iteration and 
depends solely on $\zeta$. 
In practice we can now compute $v_{\star},k_{\star},\phi_{\star}- \rho_{\star}/\rho_0\phi_0$ and $\rho_{\star}/\rho_0$ at fixed $\zeta$ as defined by the user. Once the solution of the RS equations is obtained one can use this information to correct the estimators obtained in an actual regression. For instance, to remove the overfitting bias from $\hat{\phi}_n,\hat{\sigma}_n$ and $\hat{\bbeta}_n$, one need only correct the estimators as 
\begin{eqnarray}
    \tilde{\phi}_n &=& (\hat{\phi}_n -\Tilde{\phi}_{\star})/k_{\star} = \Big(\hat{\phi}_n -\phi_{\star}- \rho_{\star}/\rho_0\phi_0\Big)/k_{\star}\\
    \tilde{\rho}_n &=& \hat{\rho}_n/k_{\star}\\
    \tilde{\bbeta}_n &=& \hat{\bbeta}_n/k_{\star}.
\end{eqnarray}
\begin{figure}[H]
\begin{subfigure}{.47\textwidth}
  \includegraphics[width=\linewidth]{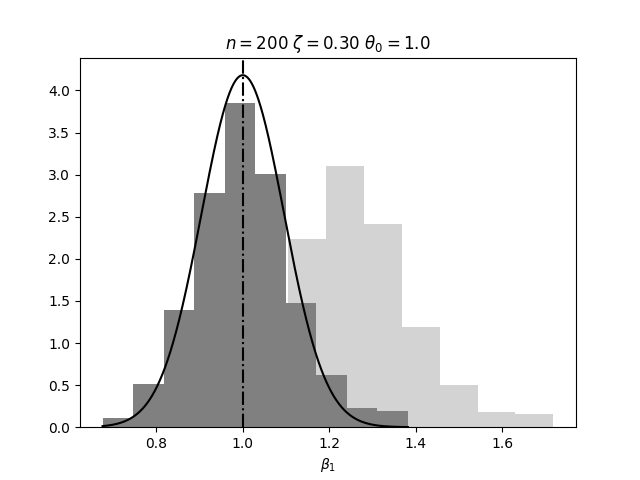}
  \caption{}
  \label{fig:3a}
\end{subfigure}%
\hfill
\begin{subfigure}{.47\textwidth}
  \includegraphics[width=\linewidth]{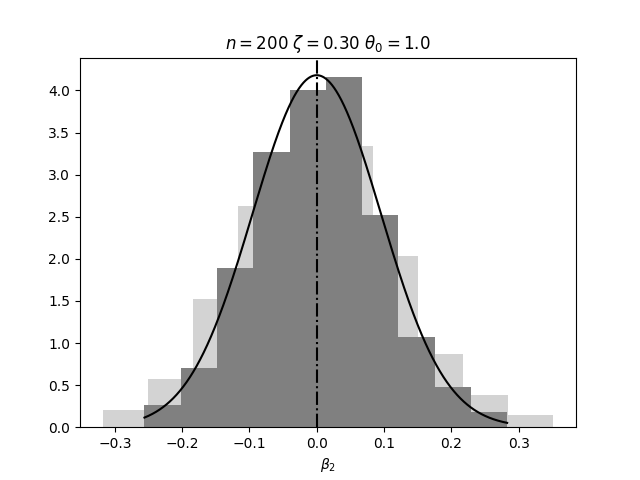}
  \caption{}
  \label{fig:3b}
\end{subfigure}
\medskip
\begin{subfigure}{.47\textwidth}
  \includegraphics[width=\linewidth]{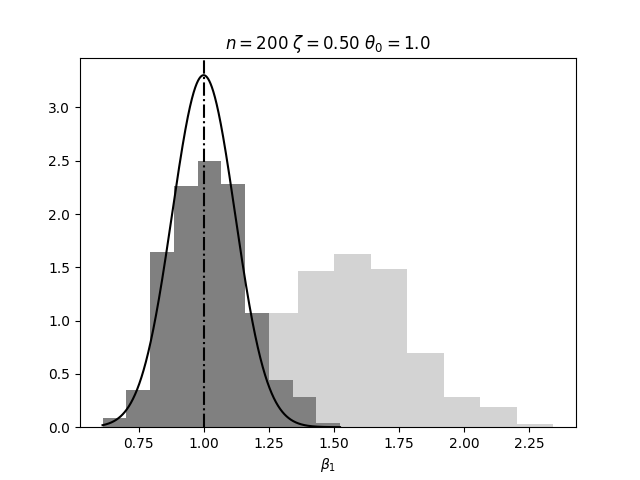}
  \caption{}
  \label{fig:3c}
\end{subfigure}%
\hfill
\begin{subfigure}{.47\textwidth}
  \includegraphics[width=\linewidth]{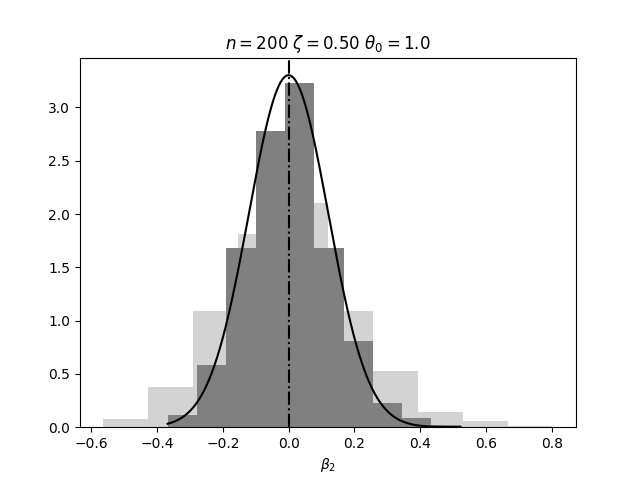}
  \caption{}
  \label{fig:3d}
\end{subfigure}
\caption{Simulated data for the Weibull model (\ref{weib_model}). Histograms based on $m=500$ realizations of the first (left) and second (right) components of the estimators $\hat{\bbeta}_n$ (light grey) and $\tilde{\bbeta}_n$ (dark grey), each based on $n=200$ i.i.d.\ samples, for two different values of $\zeta = 0.3$  (\ref{fig:3a},\ref{fig:3b}) and $\zeta =0.5$  (\ref{fig:3c},\ref{fig:3d}). Solid black line: approximate asymptotic distribution  of the de-biased estimator according to (\ref{asymp_representation}) $\tilde{\bbeta}_n \approx \mathcal{N}(\bbeta_0, v_{\star}^2/k_{\star}^2p)$. Dashed line: true value of the parameter, $\bbeta_0 = \mathbf{e}_1$. It can be seen clearly that: 1) the histogram of $\mathbf{e}_2'\tilde{\bbeta}_n$ is well described by the solid line 2) the histogram of $\mathbf{e}_1'\tilde{\bbeta}_n$ is centered at the true value, as desired, and 3)  $\mathbb{V}[\mathbf{e}_1'\hat{\bbeta}_n]=\mathbb{V}[\bbeta_0'\hat{\bbeta}_n/\|\bbeta_0\|^2]$ is underestimated.}
\label{fig:beta_weibull}
\end{figure}

In Figure \ref{fig:beta_weibull} we show the histograms of $m=500$ realizations of $\hat{\bbeta}_n$ and its corrected version $\tilde{\bbeta}_n$, as explained above. The true values are $\bbeta_0 =\mathbf{e}_1$, $\phi_0 = -\log(3)$ and $\rho_0 = 1/2$.
We note a nice agreement between the histogram of $\mathbf{e}_3'\tilde{\bbeta}_n$, which is actually null, i.e.\ $\mathbf{e}_3'\bbeta_0=0$, and the approximate asymptotic distribution $\mathcal{N}(0,v_{\star}^2/k_{\star}^2p)$.
It is noteworthy that the corrected estimator $\tilde{\bbeta}_n$ is approximately unbiased, as desired. 
Although it is also clear that the same approximation tends to underestimate the variance of the histogram of $\mathbf{e}_1'\tilde{\bbeta}_n$. All this phenomenology is in agreement with the representation (\ref{asymp_representation}). 

\begin{figure}[H]
\begin{subfigure}{.47\textwidth}
  \includegraphics[width=\linewidth]{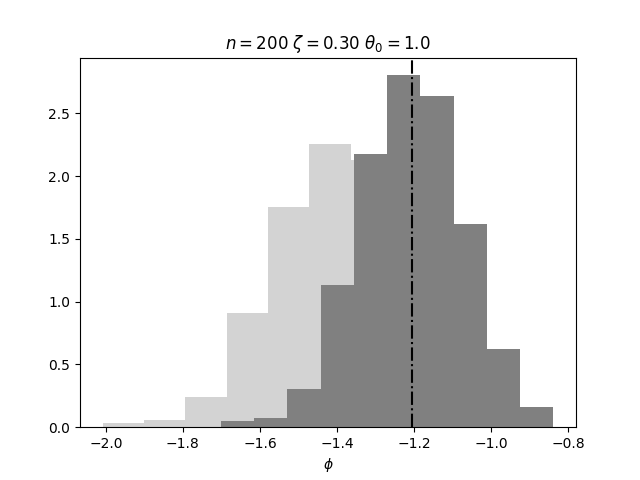}
  \caption{}
  \label{fig:4a}
\end{subfigure}%
\hfill
\begin{subfigure}{.47\textwidth}
  \includegraphics[width=\linewidth]{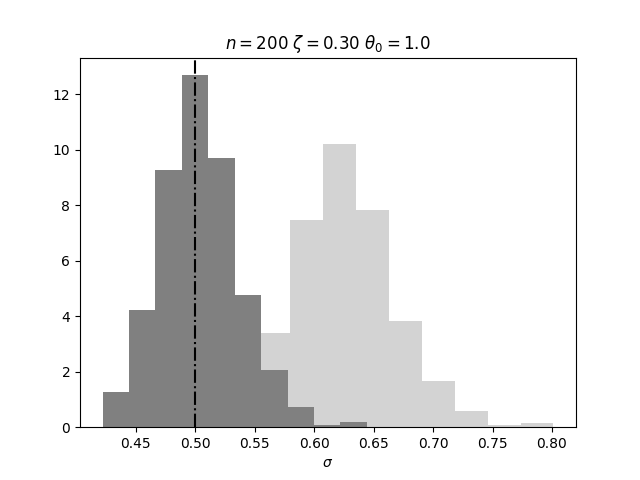}
  \caption{}
  \label{fig:4b}
\end{subfigure}
\medskip
\begin{subfigure}{.47\textwidth}
  \includegraphics[width=\linewidth]{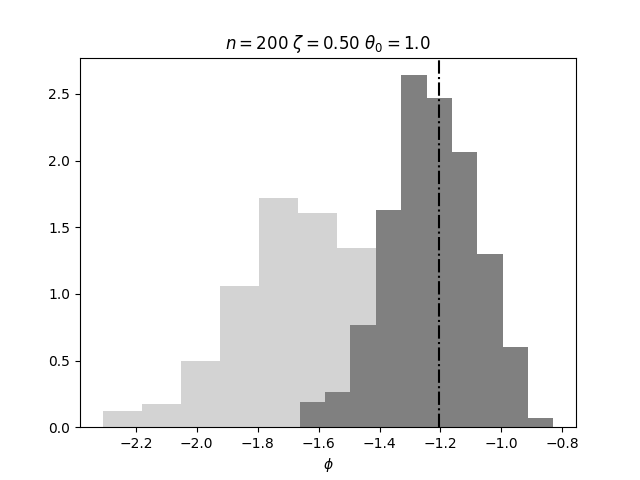}
  \caption{}
  \label{fig:4c}
\end{subfigure}%
\hfill
\begin{subfigure}{.47\textwidth}
  \includegraphics[width=\linewidth]{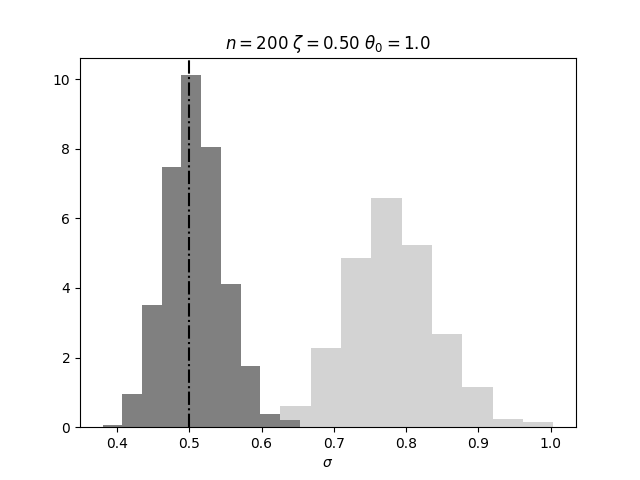}
  \caption{}
  \label{fig:4d}
\end{subfigure}
\caption{Simulated data for the Weibull model (\ref{weib_model}). The histograms of $m=500$ realizations of $\hat{\phi}_n$ (left, light grey) and $\hat{\rho}_n$ (right, light grey) together with their corrected versions $\tilde{\phi}_n$ (left, dark grey)  and $\tilde{\rho}_n$ (right, dark grey), each based on $n=200$ i.i.d. samples, for two different values of $\zeta = 0.3$  (\ref{fig:4a},\ref{fig:4b}) and $\zeta =0.5$  (\ref{fig:4c},\ref{fig:4d}).  Dashed line: true value of the parameters $\phi_0=-\log 3$, $\rho_0 = 0.5$ respectively. 
It is clear that the overfitting is already dangerous at modest values of $\zeta$. In particular $\hat{\phi}_n, \hat{\rho}_n$ are seen to clearly over-estimate $\phi_0,\rho_0$, while their corrected counterparts $\tilde{\phi}_n,\tilde{\rho}_n$ are peaked around the true values.}
\label{fig:nuis_weibull}
\end{figure}

\subsection{Logit model}
\label{subsection:logit}
The Logit regression model can be symbolically written as 
\begin{equation}
\label{logit_model}
    T = {\rm sign}\Big(\mathbf{X}'\bbeta_0+ \phi_0 + \frac{1}{2} Z\Big), \quad Z \sim \frac{\rme^{-z}}{\big(1+\rme^{-z}\big)^2}.
\end{equation}
The  log-likelihood function associated  with this model is 
\begin{equation}
\label{utility}
    u(\mathbf{X}'\bbeta,T,\phi) = - T(\mathbf{X}'\bbeta+ \phi) - \log \cosh(\mathbf{X}'\bbeta+ \phi) -\log 2.
\end{equation}
The proximal mapping ${\rm prox}_{-\tau u(.,T,\phi)}(.)$ of the minus log-likelihood, 
when viewed as a function of the linear predictor $\mathbf{X}'\bbeta$, 
satisfies
\begin{equation}
    {\rm prox}_{-\tau u(.,T,\phi)}(x) - x  + \tau T + \tau\tanh({\rm prox}_{-\tau u(.,T,\phi)}(x)+\phi)= 0.
\end{equation}
Setting $\chi(x+\phi,T) = {\rm prox}_{-\tau u(.,T,\phi)}(x)+\phi$, we get a self consistent equation defining $\chi(x+\phi,T)$
\begin{equation}
    \chi(x+\phi,T) =  x + \phi- \tau T - \tau \tanh\big(\chi(x+\phi,T)\big).
\end{equation}
After some algebraic manipulations that can be found in  Appendix \ref{appendix:logit}, we obtain that the RS equations for this model read
\begin{eqnarray}
    \label{logit1}
    v^2 \zeta &=&  \tau^2 \mathbb{E}_{T,Z_0,Q}\Big[\Big( T + \tanh(\chi_{\star})\Big)^2\Big]\\
    \label{logit2}
    \zeta &=& \mathbb{E}_{T,Z_0,Q}\Bigg[ \frac{\tau}{\cosh^2(\chi_{\star}) + \tau}\Bigg]\\
    \label{logit3}
    k\theta_0 &=& \mathbb{E}_{T,Z_0,Q}\Bigg[Z_0\chi_{\star}\Bigg]\\
    \label{logit4}
    \phi &=& \mathbb{E}_{T,Z_0,Q}\Big[\chi_{\star}\Big]  
\end{eqnarray}
where
\begin{equation}
    \chi_{\star} := \chi\big(k\theta_0Z_0 + vQ +\phi,T\big)
\end{equation}
The form of the system of equations above suggests to use fixed point iteration, and regard $\tau$ as the control parameter, rather than $\zeta$. In practice, fixing $\tau$ will be equivalent to fixing $\zeta$, when $\phi_0$ and $\theta_0$ are held fixed.

However, in applications we seldom have access to $\phi_0$ and $\theta_0 = \|\bm{\Sigma}^{1/2}_0\bbeta_0\|$, even in the ideal case of a perfectly specified model. Hence the fact that the RS equations depend on $\phi_0,\theta_0$ represents a practical problem and we have to somehow estimate these quantities from the data. An additional complication is that we only have access to estimates at fixed $\zeta$, rather than fixed $\tau$. So we must in turn convert the RS equations into a format where $\zeta$ is the control parameter.

To estimate $\theta_0$ at fixed $\zeta$, we adopt the following methodology.
Define the estimator 
\begin{eqnarray}
    \hat{\theta}_n:=\frac{1}{n}\sum_{i=1}^n \Big(\mathbf{X}_i'\hat{\bm{\beta}}_n\Big)^2 &\simeq&  \frac{1}{n}\sum_{i=1}^n \Big({\rm prox}_{-\tau u(.,T_i,\phi)}(\mathbf{X}_i'\hat{\bm{\beta}}_{(i)})\Big)^2 \\
    &=& \mathbb{E}\Big[{\rm prox}^2_{-\tau u(.,T,\phi)}(k_{\star}\theta_0Z_0 + v_{\star} Q)\Big]
\end{eqnarray}
then using the RS equations we obtain
\begin{equation}
    \mathbb{E}\Big[{\rm prox}^2_{-\tau u(.,T,\phi)}(k_{\star}\theta_0Z_0 + v_{\star} Q)\Big]= k_{\star}^2 \theta_0^2 + v_{\star}^2(1-\zeta).
\end{equation}
So we obtained an approximate equation for $\theta_0$, which we solve to estimate the latter self consistently 
\begin{equation}
\label{est_theta_0}
    \Tilde{\theta}_n = \sqrt{\big(\hat{\theta}_n -v_{\star}^2(1-\zeta)\big)/k_{\star}^2}.
\end{equation}
This amounts to adding the equation above (\ref{est_theta_0}) to the RS equations and using everywhere $\Tilde{\theta}_n$ in stead of $\theta_0$. The procedure to estimate $\phi_0$ consists in inverting the equation 
\begin{equation}
\label{implicit_phi}
    \mathbb{E}_T\Big[T\Big] + \mathbb{E}_{T,Z_0,Q}\Big[\tanh(\chi_{\star})\Big] = 0 
\end{equation}
which is equivalent to (\ref{logit4}), at fixed $\phi_{\star}$. Since $\phi_{\star}$ is not observed, we estimate it by $\hat{\phi}_n$, as they will be \say{close} for $n$ large enough. 
To see that (\ref{implicit_phi}) is an implicit equation for $\phi_0$ at fixed $\theta_0$, it is sufficient to note that
\begin{equation}
    \mathbb{E}_T\Big[T\Big]  =- \mathbb{E}_{Z_0}\Big[\tanh\big(\theta_0 Z_0 + \phi_0\big)\Big] 
\end{equation}
then 
\begin{equation}
    \mathbb{E}_{Z_0}\Big[\tanh\big(\theta_0 Z_0 + \phi_0\big)\Big] = \mathbb{E}_{T,Z_0,Q}\Big[\tanh(\chi_{\star})\Big]
\end{equation}
and we numerically invert this equation by Newton method. 
Since in an actual experiment both $\hat{\phi}_n$ and $\hat{\theta}_n$ are obtained at fixed $\zeta$, we need first to express $\zeta$ as a function of $\tau$.
This can be performed easily without the need to explicitly invert any of the RS equations. Just observe that
\begin{equation}
     \zeta = \mathbb{E}\Bigg[ \frac{\tau}{\cosh^2(\chi_{\star}) + \tau}\Bigg]
\end{equation}
can be equivalently re-written as 
\begin{equation}
    \tau = \zeta\bigg(\mathbb{E}\Bigg[ \frac{1}{\cosh^2(\chi_{\star}) + \tau}\Bigg]\bigg)^{-1}.
\end{equation}

\begin{figure}[H]
\begin{subfigure}{.47\textwidth}
  \includegraphics[width=\linewidth]{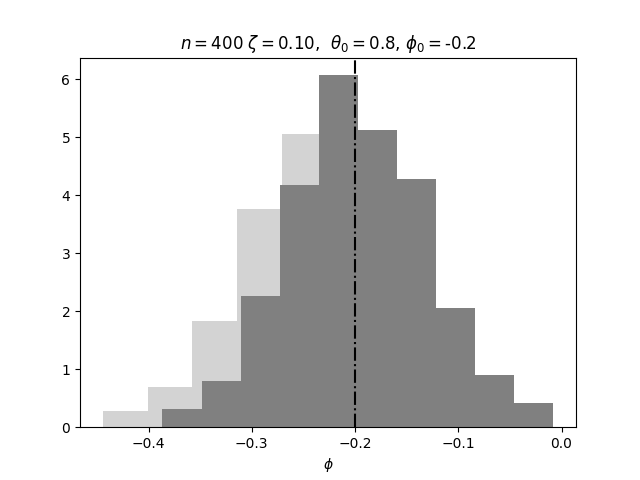}
  \caption{}
  \label{fig:6a}
\end{subfigure}%
\hfill
\begin{subfigure}{.47\textwidth}
  \includegraphics[width=\linewidth]{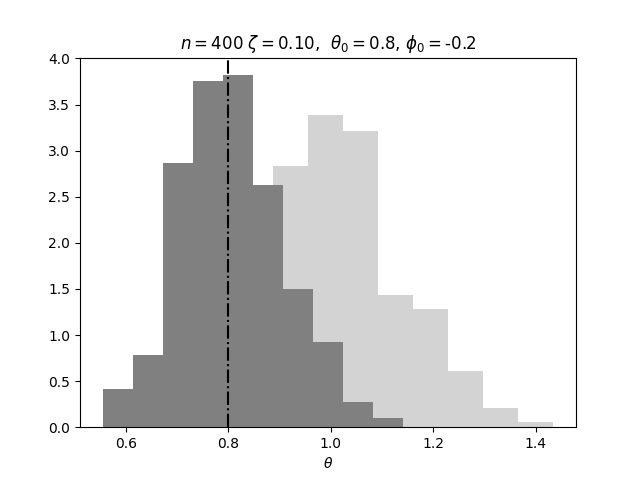}
  \caption{}
  \label{fig:6b}
\end{subfigure}
\medskip
\begin{subfigure}{.47\textwidth}
  \includegraphics[width=\linewidth]{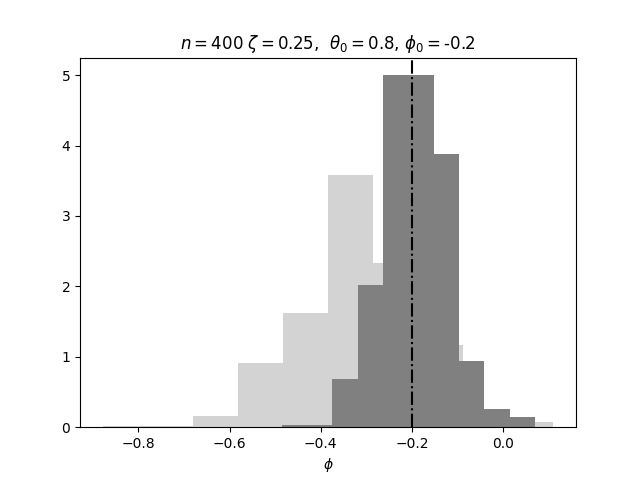}
  \caption{}
  \label{fig:6c}
\end{subfigure}%
\hfill
\begin{subfigure}{.47\textwidth}
  \includegraphics[width=\linewidth]{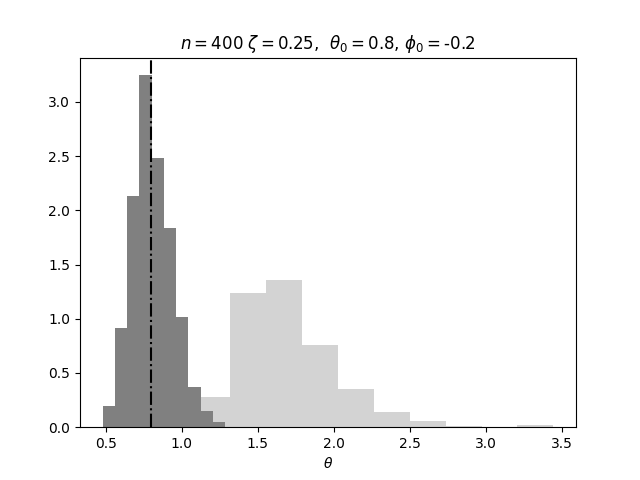}
  \caption{}
  \label{fig:6d}
\end{subfigure}
\caption{Simulated data for the Logit model (\ref{logit_model}). The histograms of $m=500$ realizations of $\hat{\phi}_n$ (left, light grey) and $\hat{\theta}_n$ (right, light grey) together with their corrected versions $\tilde{\phi}_n$ (left, dark grey) and $\tilde{\theta}_n$ (right, dark grey), each based on $n=400$ i.i.d.\ samples, for two different values of $\zeta = 0.3$  (\ref{fig:6a},\ref{fig:6b}) and $\zeta =0.5$  (\ref{fig:6c},\ref{fig:6d}). Dashed line: true value of the parameters $\phi_0=-0.2$, $\theta_0 = 0.8$.
It is clear that the overfitting is already dangerous at modest values of $\zeta$. In particular $\hat{\phi}_n,\hat{\theta}_n$ are seen to clearly over-estimate the modulus of $\phi_0, \theta_0$, respectively, already at $\zeta = 0.1$.}
\label{fig:nuis_logit}
\end{figure}
In conclusion, we have obtained a different set of RS equations 
\begin{eqnarray}
    \tilde{\theta}_n &=& \sqrt{\big(\hat{\theta}_n -v_{\star}^2(1-\zeta)\big)/k_{\star}^2}\\
    v^2 \zeta &=&  \tau^2 \mathbb{E}\Big[\Big( T + \tanh(\chi_{\star})\Big)^2\Big]\\
    \tau &=& \zeta\bigg(\mathbb{E}\Bigg[ \frac{1}{\cosh^2(\chi_{\star}) + \tau}\Bigg]\bigg)^{-1}\\
    k \hat{\theta}_0 &=& \mathbb{E}\Big[Z_0\chi_{\star}\Big]\\
    \mathbb{E}\Big[\tanh(\chi_{\star})\Big] &=&\mathbb{E}_{Z_0}\Big[\tanh\big(\theta_0 Z_0 + \phi_0\big)\Big]
\end{eqnarray}
where 
\begin{equation}
    T|Z_0\sim \frac{\rme^{-T (\tilde{\theta}_n Z_0 + \tilde{\phi}_n)}}{2\cosh(\tilde{\theta}_n Z_0 + \tilde{\phi}_n)}
\end{equation}
and 
\begin{equation}
    \chi_{\star}  =  vQ+k_{\star}\tilde{\theta}_n Z_0 + \hat{\phi}_n- \tau T - \tau \tanh(\chi_{\star}). 
\end{equation}
These equations depend only on  the quantitites $\zeta$, $\hat{\phi}_n$ and $\hat{\theta}_n$, which are indeed available in an actual regression experiment.
The solution of this set of self consistent equations can be easily obtained by fixed point iteration.

It is well known that the ML estimator $\hat{\bbeta}_n$ of the Logit regression model undergoes an asymptotically sharp phase transition, i.e. it exists with probability one if $\zeta<\zeta_c(\phi_0^2+\theta_0^2)$ and it does not exists with the same probability for $\zeta>\zeta_c$ (see \cite{HD_logit} for a detailed discussion). Hence we expect to encounter convergence issues as we approach the critical value $\zeta_c$ for finite $n,p$. We found that in our numerical simulations convergence issues arise indeed already at values of $\zeta$ much smaller than the critical value $\zeta_c$ that we compute according to \cite{HD_logit}. 
\begin{figure}[H]
\begin{subfigure}{.47\textwidth}
  \includegraphics[width=\linewidth]{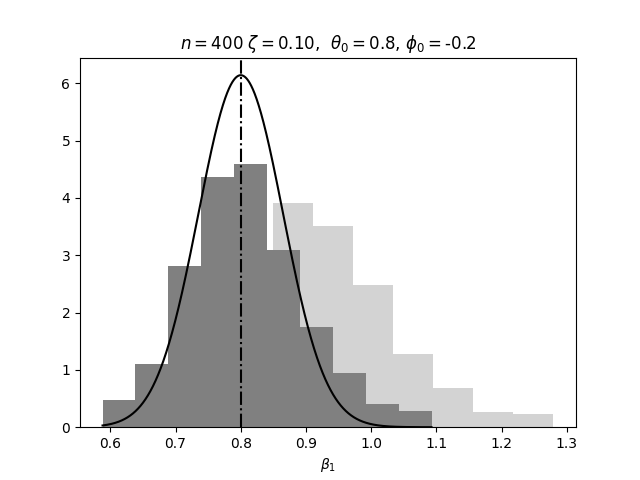}
  \caption{}
  \label{fig:5a}
\end{subfigure}%
\hfill
\begin{subfigure}{.47\textwidth}
  \includegraphics[width=\linewidth]{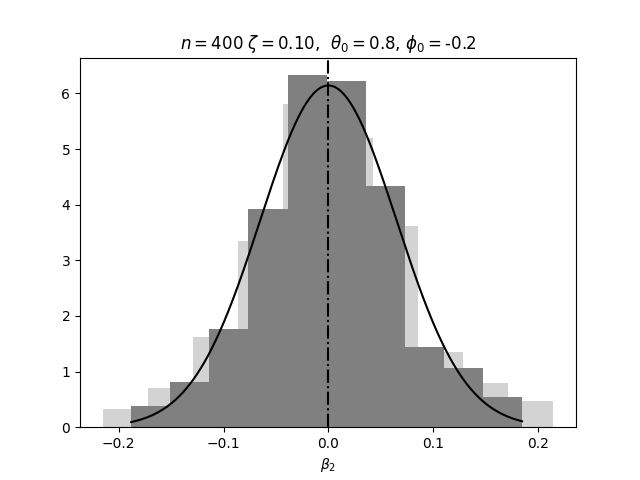}
  \caption{}
  \label{fig:5b}
\end{subfigure}
\medskip
\begin{subfigure}{.47\textwidth}
  \includegraphics[width=\linewidth]{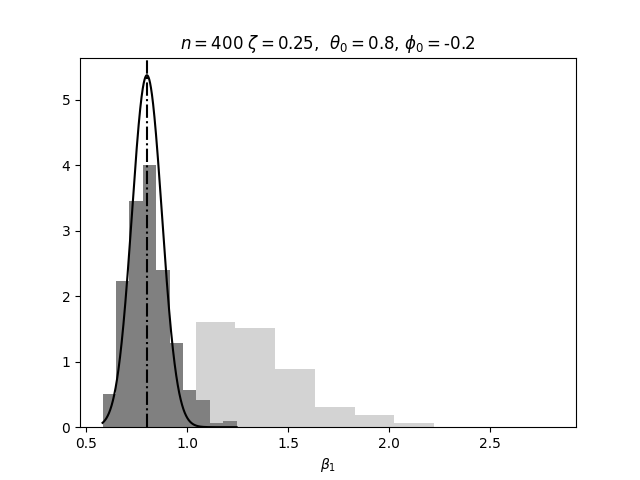}
  \caption{}
  \label{fig:5c}
\end{subfigure}%
\hfill
\begin{subfigure}{.47\textwidth}
  \includegraphics[width=\linewidth]{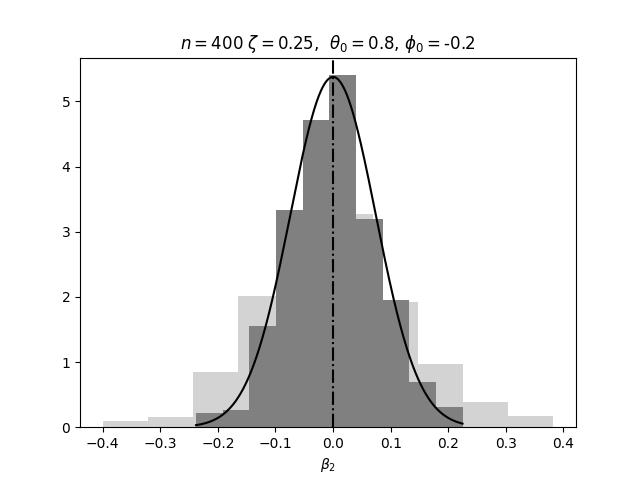}
  \caption{}
  \label{fig:5d}
\end{subfigure}
\caption{Simulated data for the Logit model (\ref{logit_model}). Histograms based on $m=500$ realizations of the first (left) and second (right) components of the estimators $\hat{\bbeta}_n$ (light grey) and $\tilde{\bbeta}_n$ (dark grey), each based on $n=400$ i.i.d. samples, for two different values of $\zeta = 0.3$  (\ref{fig:5a},\ref{fig:5b}) and $\zeta =0.5$  (\ref{fig:5c},\ref{fig:5d}). Solid black line: approximate asymptotic distribution $\mathcal{N}(\bbeta_0, v_{\star}^2/k_{\star}^2p)$ of $\tilde{\bbeta}_n$. Dashed line: true value of the parameter, $\bbeta = 0.8\mathbf{e}_1$. It can be seen clearly that: 1) the histogram of $\mathbf{e}_2'\tilde{\bbeta}_n$ is well described by the solid line 2) the histogram of $\mathbf{e}_1'\tilde{\bbeta}_n$ is centered at the true value, as wanted, 3)  $\mathbb{V}[\mathbf{e}_1'\hat{\bbeta}_n]=\mathbb{V}[\bbeta_0'\hat{\bbeta}_n/\|\bbeta_0\|^2]$ is underestimated.}
\label{fig:beta_logit}
\end{figure}

\section{Discussion and conclusion}
\label{section:discussion}
In this article we have extended the existing literature on regression with generalized linear models in the proportional asymptotic regime ($p = \zeta n$). 
We showed via a novel route how to deduce the asymptotic properties of the ML estimators of the regression parameters and the nuisance parameters in the model from a small set of equations. 
The approximate asymptotic distribution of the estimator $\hat{\bbeta}_n$ for the regression parameters is described by two non random values $k_{\star}$ and $v_{\star}$ that can be computed by solving these equations. At fixed $p,n$ and hence $\zeta$, $k_{\star}$ and $v_{\star}$ have a precise geometric interpretation via the representation introduced in \cite{massa} and the hypothesis of concentration of the overlaps, as explained in the main text. Their values can be used to: 1) obtain an approximate asymptotic distribution of the components of $\hat{\bbeta}_n$ and 2) to compute correction factors to simultaneously reduce the bias in $\hat{\bbeta}_n$ and the inflation of the variance of $\hat{\bbeta}_n$, which are caused by the high dimension of the model ($p$), compared to the sample size ($n$). The estimators of the nuisance parameters $\hat{\bsigma}_n$ are also affected by bias, and the solution of the self consistent equations provides a way to \say{de-bias} the estimates. We studied via simulations the ML estimators and their corrected counterparts, when the true regression parameter $\bbeta_0$ is sparse. We show that the corrected estimators are effectively un-biased. Furthermore, we find that in general the approximate asymptotic representation (\ref{asymp_representation}) is accurate in describing: 1) the distribution of the null components of $\hat{\bbeta}_n$ (i.e.\ those for which the true value is zero), 2) the mode of the histogram of $\hat{\bsigma}_n$. Contrary to the case where $\bbeta_0$ is diffuse \cite{GLM,sheikh,HD_logit}, we see that when $\mathbf{e}_k'\bbeta_0=O_n(1)$, $\mathbb{V}[\mathbf{e}_k'\hat{\bbeta}_n]$ is larger than the value $v_{\star}/\sqrt{p}$ obtained from the theory. At the moment, confidence statement are only possible for $\hat{\bbeta}_n$, as we do not have access to any information on the finite size fluctuations (i.e.\ the variance) of $\hat{\bsigma}_n$. It is necessary to obtain more information, even if approximate, on the distribution of $\hat{\bsigma}_n$ in order to allow confidence statements on the estimate. This is likely to be obtained from a generalization of the current theory and will be subject of future investigations. 


Our results are derived, for analytical convenience, under the assumption that $\mathbf{X}\sim \mathcal{N}(\bm{0}, \bm{\Sigma}_0)$.  Rotational invariance of the density of $\mathbf{X}$ is the only truly necessary ingredient to reach the stochastic representation (\ref{representation_beta}). Furthermore, even if the representation might not hold for other distributions, the asymptotic bias, together with the variance of the null components would still be given by the RS equations (\ref{rs1}, \ref{rs2}, \ref{rs3}, \ref{rs4}) provided a central limit theorem applies to $\mathbf{X}_i'\hat{\bbeta}_{(i)}$, conditional on $\mathbf{X}_{(i)} = (X_1,\dots,X_{i-1},X_{i+1},\dots,X_p)'$. Hence we expect that our result should be of interest also for cases which do not strictly fall in the setting of the present paper.

The Python implementations of the routines used in section (\ref{section:applications}) are available at 
https://github.com/EmanueleMassa/Correction\_of\_overfitting\_bias . These numerical routines take as input only observable quantities like $\zeta$, the ML estimators $\hat{\bbeta}_n,\hat{\bsigma}_n$ and eventually the data $\{(T_i,\mathbf{X}_i)_{i=1}^n\}$ in order to return correction factors for the estimators as explained in the application section.


\begin{appendix}
\section{Representation of $\hat{\bbeta}_n$}
\label{appendix:representation}

We denote equality in distribution with $\overset{d}{=}$. Since for $\mathbf{X}\sim \mathcal{N}(\bm{0},\bm{\Sigma}_0)$
\begin{equation}
    \mathbf{X} \overset{d}{=}\bm{\Sigma}_0^{1/2} \bm{\mathcal{X}}, \quad \bm{\mathcal{X}}\sim \mathcal{N}(\bm{0},\bm{I}),
\end{equation}
it follows that 
\begin{equation}
    \hat{\bbeta}_n\big(\bbeta_0,\{(T_i,\mathbf{X}_i)\}\big) \overset{d}{=} \bm{\Sigma}_0^{-1/2}\tilde{\bbeta}_n\big(\Tilde{\bbeta}_0,\{(T_i,\bm{\mathcal{X}}_i)\}\big)
\end{equation}
with 
\begin{equation}
    \tilde{\bbeta}_n\big(\Tilde{\bbeta}_0,\{(T_i,\bm{\mathcal{X}}_i\}\big)=  \underset{\bbeta}{\arg\max}\bigg( \underset{\bsigma}{\max}\Big\{ \sum_{i=1}^n u (\bm{\mathcal{X}}_i'\bbeta,T_i,\bsigma) \Big\}\bigg)
\end{equation}
where 
\begin{equation}
    T_i|\bm{\mathcal{X}}_i \sim f(.|\bm{\mathcal{X}}'\Tilde{\bbeta}_0,\bsigma_0), \quad  \bm{\mathcal{X}}_i := \bm{\Sigma}_0^{-1/2}\mathbf{X}_i \sim  \bm{\mathcal{N}}(\bm{0},\bm{I})
\end{equation}
with $\Tilde{\bbeta}_0:= \bm{\Sigma}_0^{1/2}\bbeta_0$. Now consider any rotation $\mathbf{R}_0$ in $\mathbb{R}^p$ around $\Tilde{\bbeta}_0$. Such rotations applied to all $\bm{\mathcal{X}}_i$ leave both the distribution of each $\bm{\mathcal{X}}_i$ and the value of each  $\bm{\mathcal{X}}_i'\Tilde{\bbeta}_0$ invariant. 
Hence, since $T_i|\bm{\mathcal{X}}_i  = T_i|\bm{\mathcal{X}}_i'\Tilde{\bbeta}_0$, the joint distribution  of each pair $(\bm{\mathcal{X}}_i,T_i)$ is invariant.  
We therefore obtain
\begin{align}
\label{invariance}
     \Tilde{\bbeta}_n (\{(T_i,\bm{\mathcal{X}}_i)\},\Tilde{\bbeta}_0)& \overset{d}{=}~ \mathbf{R}_0 \Tilde{\bbeta}_n (\{(T_i,\mathbf{R}_0\bm{\mathcal{X}}_i)\},\Tilde{\bbeta}_0) \nonumber\\
    &\overset{d}{=}~ \mathbf{R}_0 \Tilde{\bbeta}_n(\{(T_i,\bm{\mathcal{X}}_i)\},\Tilde{\bbeta}_0)
\end{align} 
To understand the implications of (\ref{invariance}),  we  decompose $\tilde{\bbeta}_n:=\tilde{\bbeta}_n\{(T_i,\bm{\mathcal{X}}_i)\},\Tilde{\bbeta}_0)$ into a component along $\Tilde{\bbeta}_0$ and a component in the subspace orthogonal to $\Tilde{\bbeta}_0$
\begin{equation}
    \Tilde{\bbeta}_n  =  \Tilde{\bbeta}_{n,\|} +  \Tilde{\bbeta}_{n,\perp} \qquad \Tilde{\bbeta}_{n,\|}' \Tilde{\bbeta}_{n,\perp}=0
\end{equation}
with 
\begin{equation}
    \Tilde{\bbeta}_{n,\|}:=\Tilde{\bbeta}_0(\Tilde{\bbeta}_0'\Tilde{\bbeta}_n)/\|\Tilde{\bbeta}_0\|^2 \qquad \Tilde{\bbeta}_{n,\perp} := \Big(\bm{I}-\Tilde{\bbeta}_0\Tilde{\bbeta}_0'/\|\Tilde{\bbeta}_0\|^2\Big)\Tilde{\bbeta}_n.
\end{equation}
Then equation (\ref{invariance}) implies that 
\begin{equation}
    \mathbf{R}_0\Tilde{\bbeta}_{n,\perp} \overset{d}{=} \Tilde{\bbeta}_{n,\perp}
\end{equation}
 Hence all the values of $\Tilde{\bbeta}_{n,\perp}$ that have the same length, i.e.\ that lie on a sphere in the subspace of $\mathbb{R}^p$ orthogonal to $\Tilde{\bbeta}_0$, must have the same probability density.
Conditional on its length, the direction of $\Tilde{\bbeta}_{n,\perp}$ is uniformly distributed over a sphere in the above subspace.
This means that, conditional on $\Tilde{\bbeta}_{n,\|}$, we have
\begin{equation}
    \Tilde{\bbeta}_{n,\perp} =  \|\Tilde{\bbeta}_{n,\perp}\| \mathbf{U},~~~~~
\mathbf{U}\sim {\rm Unif}(\mathcal{S}_{p-2}),~ \mathbf{U}'\Tilde{\bbeta}_0=0
\end{equation}
Finally, using 
\begin{equation}
    \|\Tilde{\bbeta}_{n,\perp}\| =   \sqrt{\|\Tilde{\bbeta}_n\|^2-(\Tilde{\bbeta}_0'\Tilde{\bbeta}_n)^2/\|\Tilde{\bbeta}_0\|^2}
\end{equation}
then leads us to the following representation, in which $\mathbf{Z}\sim \mathcal{N}(\bm{0},\bm{I}_p)$: 
\begin{equation}
\label{eq:transformed_rep1}    \Tilde{\bbeta}_n = K_n \Tilde{\bbeta}_0+ V_n \mathbf{U},
    ~~~~
    \mathbf{U} = \frac{\big(\bm{I} - \Tilde{\bbeta}_0\Tilde{\bbeta}_0'/\|\Tilde{\bbeta}_0\|^2\big) \mathbf{Z}}{\Big\| \big(\bm{I} - \Tilde{\bbeta}_0\Tilde{\bbeta}_0'/\|\Tilde{\bbeta}_0\|^2\big) \mathbf{Z}\Big\|},
\end{equation}
with
\begin{eqnarray}
\label{eq:transformed_rep2}
    K_n = (\Tilde{\bbeta}_0' \Tilde{\bbeta}_n)/\|\Tilde{\bbeta}_0\|^2,~~~~~
    V^2_n = \| \Tilde{\bbeta}_n\|^2 - K_n^2\|\Tilde{\bbeta}_0\|^2.
\end{eqnarray}
Upon finally transforming (\ref{eq:transformed_rep1}, \ref{eq:transformed_rep2}) back into the original basis, via $\tilde{\bbeta}_n=\bm{\Sigma}_0^{\frac{1}{2}}\hat{\bbeta}_n$ and $\tilde{\bbeta}_0=\bm{\Sigma}_0^{\frac{1}{2}}\bbeta_0$, 
we obtain 
\begin{eqnarray}
&&\label{eq:rep1}    \hat{\bbeta}_n = K_n \bbeta_0+ V_n \bm{\Sigma}_0^{-\frac{1}{2}}\mathbf{U},~~~~~~\mathbf{U}'\bm{\Sigma}_0^{\frac{1}{2}}\bbeta_0=0
\\
\label{eq:rep2}
 &&   K_n = \frac{\bbeta_0'\bm{\Sigma}_0\hat{\bbeta}_n}{\bbeta_0'\bm{\Sigma}\bbeta_0},~~~~~
    V^2_n = \hat{\bbeta}_n'\bm{\Sigma}_0\hat{\bbeta}_n-\frac{(\bbeta_0'\bm{\Sigma}\hat{\bbeta}_n)^2}{\bbeta_0'\bm{\Sigma}_0\bbeta_0}.
\end{eqnarray}

\section{RS equations for linear models in the location-scale family }
\label{appendix:location_scale}

The particular class of the linear models 
\begin{equation}
    T = \mathbf{X}'\bbeta + \epsilon, \quad \epsilon \sim f_{\epsilon}(.|\bsigma)
\end{equation}
in the location scale family are particularly important and widely adopted. 
The distribution of $\epsilon$ is in the location-scale family if $\epsilon \sim f_{\epsilon}(.|\bsigma)$ with $\bsigma = (\phi,\sigma)$ and 
\begin{equation}
    \epsilon \overset{d}{=} \phi + \sigma Z,  \quad Z \sim f_Z
\end{equation}
where the law $f_Z$ is standard, i.e.\ $f_Z(.) = f_{\epsilon}(.|\phi = 0,\sigma=1)$.
This means that 
\begin{equation}
    f_{\epsilon}(x|\phi,\sigma) = f_Z\big(\frac{x-\phi}{\sigma}\big)/\sigma
\end{equation}
and
\begin{equation}
    \log f_{\epsilon}(\epsilon|\phi,\sigma) = \log f_Z\big(\frac{\epsilon -\phi}{\sigma}\big) - \log (\sigma/\sigma_0) - \log \sigma_0.
\end{equation}
We explained in the main text (see Subsection \ref{subsection:linear}) that for linear models we can always set $\theta_0:=|\bbeta_0| = 0$ to derive the RS equations, as these do not depend on $\theta_0$.
Hence, without loss of generality, we may take $T \overset{d}{=}\phi_0 + \sigma_0 Z$, obtaining
\begin{equation}
    u(x,T,\phi,\sigma) = \log f_Z\big(Z \frac{\sigma_0}{\sigma} + \frac{\phi_0-\phi}{\sigma}-\frac{x}{\sigma}\big)- \log (\sigma/\sigma_0) + C
\end{equation}
in which  $C$ is independent of $\phi$ and $\sigma$.
Now 
\begin{eqnarray}
    \xi &=& \underset{y}{\arg \min}\bigg\{ \frac{1}{2} (y-vQ)^2 - \tau \log f_Z\Big(\frac{\phi_0-\phi}{\sigma} + \frac{\sigma_0}{\sigma} Z - y/\sigma\Big)\bigg\} \nonumber \\
    &=&  -\sigma ~\underset{y}{\arg \min}\bigg\{ \frac{1}{2} \Big(\frac{\phi_0-\phi}{\sigma} + \frac{\sigma_0}{\sigma} Z-\frac{v}{\sigma} Q - y\Big)^2 - \frac{\tau}{\sigma^2} \log f_Z(y)\bigg\}\nonumber\\
    &~~&  ~~ + \; \sigma \big(\frac{\phi_0-\phi}{\sigma} + \frac{\sigma_0}{\sigma} Z\big)
\end{eqnarray}
It is then convenient to define 
\begin{equation}
    \Tilde{\xi}= -{\rm prox}_{-\tilde{\tau} \tilde{u}(.)}\Big(\frac{\phi_0-\phi}{\sigma} + \frac{\sigma_0}{\sigma} Z-\frac{v}{\sigma} Q
    \Big)
\end{equation}
with the short-hands $\tilde{\tau} :=\tau/\sigma^2$ and $\tilde{u}:=\log f_Z$.
We then have
\begin{equation}
\label{linear_simp1}
    \xi= \sigma \Big(\Tilde{\xi}+\frac{\phi_0-\phi}{\sigma} + \frac{\sigma_0}{\sigma} Z\Big).
\end{equation}
Substitution of (\ref{linear_simp1}) into the RS equations (\ref{rs1}, \ref{rs2}), setting $\theta_0 = 0$ and $k=1$ because the model is linear (see Subsection \ref{subsection:linear}), and with 
\begin{equation}
     \ddot{u}(\xi,T,\phi,\sigma)= \frac{1}{\sigma^2}\ddot{\tilde{u}}(-\tilde{\xi})
\end{equation}
then gives equations (\ref{lin:rs1}, \ref{lin:rs2})  of subsection \ref{subsection:linear}.
The equations for the nuisance parameters (\ref{lin:rs3},\ref{lin:rs4}) are obtained by noticing that 
\begin{eqnarray}
    \label{phi_der1}
    \frac{\partial}{\partial \phi} u(\xi,\epsilon,\phi,\sigma) &=& \frac{\partial}{\partial \phi}\log f(\epsilon-\xi|\phi,\sigma) =-\frac{1}{\sigma} \dot{\tilde{u}}(-\tilde{\xi}) = \nonumber\\
    &=& \frac{1}{\sigma \tilde{\tau}}(\Tilde{\xi}+v/\sigma Q - (\phi-\phi_0)/\sigma + \frac{\sigma_0}{\sigma} Z)\\
    \label{sigma_der1}
    \frac{\partial}{\partial \sigma} u(\xi,\epsilon,\phi,\sigma) &=& \frac{\partial}{\partial \sigma}\log f(\epsilon-\xi|\phi,\sigma)= -\frac{1}{\sigma} +\frac{1}{\sigma}
    \Tilde{\xi}\dot{\tilde{u}}(-\tilde{\xi}) 
\end{eqnarray}
where we used the definition of proximal operator (\ref{def:prox}).
Taking the expectation of (\ref{phi_der1}) and setting the result to zero gives
\begin{equation}
    (\phi-\phi_0)/\sigma = \mathbb{E}\Big[\Tilde{\xi}\Big] + (\frac{\sigma_0}{\sigma})\mathbb{E}\Big[Z\Big] 
\end{equation}
which is equation (\ref{lin:rs3}) in the main text.
Then, taking the expectation of (\ref{sigma_der1}) and setting the result to zero we get
\begin{equation}
    \mathbb{E}\Big[\Tilde{\xi}\dot{\tilde{u}}(-\Tilde{\xi})\Big]= -\mathbb{E}\Big[\Tilde{\xi}\big(\Tilde{\xi}+v/\sigma Q - (\phi-\phi_0)/\sigma + \frac{\sigma_0}{\sigma} Z\big)/\tilde{\tau}\Big] = 1 
\end{equation}
which implies, upon using (\ref{lin:rs1}) and (\ref{lin:rs2}),
\begin{eqnarray}
   -\mathbb{E}\Big[\Tilde{\xi}\dot{\tilde{u}}(-\Tilde{\xi})\Big] &=&\mathbb{E}\Big[\big(\Tilde{\xi}+v/\sigma Q - (\phi-\phi_0)/\sigma + \frac{\sigma_0}{\sigma} Z\big)^2\Big] \nonumber \\
    &~~&  ~~ - \; v/\sigma\mathbb{E}\Big[Q\big(\Tilde{\xi}+v/\sigma Q - (\phi-\phi_0)/\sigma + \frac{\sigma_0}{\sigma} Z\big)\Big]  \nonumber \\
    &~~&  ~~ - \; \frac{\sigma_0}{\sigma} \mathbb{E}\Big[Z\big(\Tilde{\xi}+v/\sigma Q - (\phi-\phi_0)/\sigma + \frac{\sigma_0}{\sigma} Z\big)\Big]  \nonumber\\
    &=&-\frac{\sigma_0}{\sigma} \mathbb{E}\Big[Z\big(\Tilde{\xi}-v/\sigma Q - (\phi-\phi_0)/\sigma - \frac{\sigma_0}{\sigma} Z\big)\Big] \nonumber \\
    &=&\tau \frac{\sigma_0}{\sigma} \mathbb{E}\Big[Z\dot{\tilde{u}}(-\Tilde{\xi})\Big].
\end{eqnarray}
Hence we obtain equation (\ref{lin:rs4}) in the main text:
\begin{equation}
    \sigma/\sigma_0 =  -\mathbb{E}[Z \dot{u}(-\Tilde{\xi}) ].
\end{equation}

\section{Leave-one-out method approximations}
 \label{appendix:leave_one_out}
 

\subsection{Approximation of the linear predictor}
\label{appendix:subsection_lp}
For fixed $\bsigma$, the value of $\bbeta$ that maximizes the log-likelihood satisfies
\begin{equation}
     \frac{1}{n} \sum_{i=1}^n \mathbf{X}_i \dot{u}(\mathbf{X}_i'\hat{\bbeta}_n,T_i,\bsigma)= 0
\end{equation}
where we used the notation $ \dot{u}(x,T,\bsigma) = \partial_{x} u(x,T,\bsigma)$.
Similarly, upon excluding  the $i$-th observation, the resulting leave-one-out  estimator $\hat{\bbeta}_{(i)}$  satisfies 
\begin{equation}
    \frac{1}{n} \sum_{j\neq i} \mathbf{X}_j \dot{u}(\mathbf{X}_j'\hat{\bbeta}_{(i)},T_j,\bsigma)= 0.
\end{equation}
We use a first order Taylor expansion around 
$\mathbf{X}_j'\hat{\bbeta}$ and after simplifying and rearranging the terms we get 
\begin{equation}
\label{beta_sur}
    \hat{\bbeta}_n-\hat{\bbeta}_{(i)} = \frac{1}{n} \mathbf{I}^{-1}_{(i)}(\bar{\bbeta}_{i})\mathbf{X}_i \dot{u}(\mathbf{X}_i'\hat{\bbeta},T_i,\bsigma)
\end{equation}
where, with $\ddot{u}(x,T_j,\bsigma) = \partial^2_{x} u(x,T,\bsigma)$,
\begin{equation}
    \mathbf{I}_{(i)}(\bar{\bbeta}_{i}):= - \frac{1}{n}\sum_{j\neq i} \mathbf{X}_j\mathbf{X}_j' \ddot{u}(\mathbf{X}_j'\bar{\bbeta}_{i},T_j,\bsigma).
\end{equation}
with 
\begin{equation}
    \bar{\bbeta}_{i} :=  s \hat{\bbeta}_n + (1-s) \hat{\bbeta}_{(i)}, \quad s \in (0,1).
\end{equation}
If we then take the scalar product of (\ref{beta_sur}) with $\mathbf{X}_i$ we obtain an equation relating the leave-one-out linear predictor to the actual one
\begin{equation}
\label{leave_one_out_eq}
    \mathbf{X}_i'(\hat{\bbeta}_n - \hat{\bbeta}_{(i)}) = \tau_i \dot{u}(\mathbf{X}_i'\hat{\bbeta}_n,T_i,\bsigma)
\end{equation}
where
\begin{equation}
\label{tau_i}
    \tau_i := \frac{1}{n}\mathbf{X}_i'\Big(\mathbf{I}_{(i)}(\bar{\bbeta}_{i})\Big)^{-1}\mathbf{X}_i.
\end{equation}
We recognize (\ref{leave_one_out_eq}) to be the equation defining the proximal mapping of \\
$-\tau_i u(.,T_i,\bsigma)$, thus 
\begin{equation}
\label{prox}
    \mathbf{X}_i'\hat{\bbeta}_n = {\rm prox}_{-\tau_i u(.,T_i,\bsigma)}(\mathbf{X}_i'\hat{\bbeta}_{(i)}),
\end{equation}
where the proximal mapping of a convex function $f:\mathbb{R}\rightarrow \mathbb{R}$ is defined as 
\begin{equation}
\label{def:prox}
     {\rm prox}_{f}(x) := \underset{y}{\arg\min}\Big\{\frac{1}{2}(y-x)^2 + f(y)\Big\}.
\end{equation}

The problem with (\ref{prox})  is that we do not know $\bar{\bbeta}_i$. On the other hand $\bar{\bbeta}_i$ will be \say{close} to $\hat{\bbeta}_{(i)}$, as  $\hat{\bbeta}_{(i)}$ is in turn expected to be close to $\hat{\bbeta}_n$.
This motivates the introduction of the following approximation for the linear predictor $\mathbf{X}_i'\hat{\bbeta}_n$
\begin{equation}
\label{xi}
    \xi_i = {\rm prox}_{-\tau_n u(.,T_i,\bsigma)}\big(\mathbf{X}_i'\hat{\bbeta}_{(i)}\big)
\end{equation}
with 
\begin{equation}
    \tau_n := \frac{1}{n} \mathbf{X}_i'\mathbf{I}_{(i)}^{-1}(\hat{\bbeta}_{(i)}) \mathbf{X}_i =  \frac{1}{n}\Tr\big(\mathbf{I}_{(i)}^{-1}(\hat{\bbeta}_{(i)})\big) + o_P(1)
\end{equation}
because of the standard concentration properties of quadratic forms of normal random vectors as $\mathbf{I}_{(i)}^{-1}(\hat{\bbeta}_{(i)}) \perp \mathbf{X}_i$.

\subsection{The equation for $\tau_n$}
\label{appendix:subsection_tau}

In the previous section we implicitly assumed that $\mathbf{I}_{(i)}(\hat{\bbeta}_{(i)})$ is invertible.
On the other hand, assuming that also $\mathbf{I}(\hat{\bbeta}_n)$ is invertible, we have 
\begin{eqnarray}
     p &=&\Tr\Big(\mathbf{I}(\hat{\bbeta}_n)\mathbf{I}^{-1}(\hat{\bbeta}_n)\Big) = -\frac{1}{n}\sum_{j=1}^n \ddot{u}(\mathbf{X}_j'\hat{\bbeta}_n,T_j,\bsigma) \mathbf{X}_j'\mathbf{I}^{-1}(\hat{\bbeta}_n)\mathbf{X}_j \nonumber \\
    &=& \sum_{i=1}^n \bigg(1-\frac{1}{1-\frac{1}{n}\ddot{u}(\mathbf{X}_i'\hat{\bbeta}_n,T_i,\bsigma)\mathbf{X}_i'\mathbf{I}_{(i)}^{-1}(\hat{\bbeta}_n)\mathbf{X}_i}\bigg)
\end{eqnarray}
where we used the Sherman-Morrison-Woodbury formula
\begin{equation}
    \mathbf{v}'(\mathbf{A} + \mathbf{v}\mathbf{v}')^{-1}\mathbf{v}=  1-\frac{1}{1+\mathbf{v}'\mathbf{A}^{-1}\mathbf{v}}
\end{equation}
with $\mathbf{v} = \sqrt{-\ddot{u}(\mathbf{X}_i'\hat{\bbeta}_n,T_i,\bsigma)}\mathbf{X}_i$  and $\mathbf{A} = \mathbf{I}_{(i)}^{-1}(\hat{\bbeta}_n)$.
If  the approximation (\ref{xi}) of $\mathbf{X}_i\hat{\bbeta}_n$ is accurate and we can approximate 
\begin{equation}
    \mathbf{X}_i'\mathbf{I}_{(i)}^{-1}(\hat{\bbeta}_n)\mathbf{X}_i \simeq \mathbf{X}_i'\mathbf{I}_{(i)}^{-1}(\hat{\bbeta}_{(i)})\mathbf{X}_i = \tau_n + o_P(1)
\end{equation}
then 
\begin{equation}
\label{fz_approx}
    p \simeq \sum_{i=1}^n \bigg(1-\frac{1}{1-\tau_n\ddot{u}(\mathbf{X}_i'\hat{\bbeta}_{n},T_i,\bsigma)}\bigg) \ .
\end{equation}
For the approximation above to be fully rigorous one would have to bound the reminders, which has been indeed done for some particular $u$ (see the excellent references \cite{el_karoui_rigorous} for robust regression and \cite{HD_logit} for logit regression). This is not easily shown for general non-linear models, so here we instead verify these assumptions a posteriori.
After some straightforward algebraic simplifications, using $\zeta = p/n$, we get 
\begin{equation}
\label{zeta_approx}
    1-\zeta  \simeq  \frac{1}{n}\sum_{i=1}^n \frac{1}{1-\tau_n\ddot{u}(\mathbf{X}_i'\hat{\bbeta}_n,T_i,\bsigma) }.
\end{equation}
Note that using 
\begin{equation}
    \frac{\rmd }{\rmd x} {\rm prox}_{f} (x) = \dot{{\rm prox}}_{f} (x) = \frac{1}{1+ \ddot{f}\Big({\rm prox}_{f} (x)\Big)}
\end{equation}
we can also re-write (\ref{fz_approx}) as 
\begin{equation}
\label{eq_u}
    1-\zeta = \frac{1}{n}\sum_{i=1}^n\dot{{\rm prox}}_{-\tau_n u(.,T_i,\bsigma)}(\mathbf{X}_i'\hat{\bbeta}_{n},T_i,\bsigma).
\end{equation}

\subsection{Self consistent equations } 
\label{appendix:subsection_self_consistent_eqs}
As a consequence of (\ref{eq_k}, \ref{eq_v}, \ref{eq_tau}, \ref{eq_sigma}) in the main text, we established the following set of equations 
\begin{eqnarray}
    &&\mathbb{E}\Big[ Z_0\Big(\xi - k\theta_0 Z_{0}  -vQ\Big)/\tau\Big] = 0 \\
    &&\mathbb{E}\Big[ \xi \dot{u}\Big(\xi,T,\bsigma)\Big)\Big] = 0 \\
    \label{eq_sigma_final}
    && \mathbb{E}\Big[\mathbf{g}\big(\xi,T,\bsigma \big)\Big] = 0\\
    \label{eq_u_final}
    &&\mathbb{E}\Big[\frac{1}{1-\tau \ddot{u}(\xi,T,\bsigma) } \Big] = \tau.
\end{eqnarray}
First of all note that (\ref{eq_sigma_final}, \ref{eq_u_final}) are exactly the equations (\ref{rs4}, \ref{rs2}) in the main text. Then, since $Z_0\perp Q$ and $\mathbb{E}[Z_0^2] = 1$, we get
\begin{equation}
\label{eq_k_final}
    \mathbb{E}\Big[ Z_0\Big(\xi - k\theta_0 Z_{0}  -vQ\Big)\Big] =  \tau \mathbb{E}\Big[ Z_0\dot{u}(\xi,T,\bsigma)\Big] = 0 \ \implies \ k\theta_0 =  \mathbb{E}\Big[ Z_0\xi\Big]
\end{equation}
which is equation (\ref{rs3}) in the main text.
Now observe that by (\ref{eq_u}) and integration by parts we have 
\begin{equation}
\label{eq_tau_final}
    \tau= \mathbb{E}\Big[\dot{{\rm prox}}_{-\tau u(.,T,\bsigma)}(\xi,T,\bsigma)\Big]  = \mathbb{E}\Big[Q \xi\Big].
\end{equation}
and also 
\begin{eqnarray}
    &&\hspace*{-9mm}\mathbb{E}\Big[ \xi^2\Big] -\mathbb{E}\Big[ (k\theta_0Z_0+vQ)\xi\Big]  \\
    &&=\mathbb{E}\Big[ \Big(\xi-(k\theta_0Z_0+vQ)\Big)^2 \Big] + \mathbb{E}\Big[ \Big(\xi-(k\theta_0Z_0+vQ)\Big)(k\theta_0Z_0+vQ)\Big]. \nonumber 
\end{eqnarray}
hence by (\ref{eq_k_final}, \ref{eq_tau_final}) we conclude that 
\begin{equation}
    \mathbb{E}\Big[ \Big(\xi-(k\theta_0Z_0+vQ)\Big)^2 \Big] = \tau^2 \mathbb{E}\Big[ \Big(\dot{u}(\xi,T,\bsigma)\Big)^2 \Big] =  v^2 \zeta .
\end{equation}
which is equation (\ref{rs1}) in the main text.

\section{Details for the Log-logistic AFT model}
\label{appendix:log_log}
Substitution of $\Tilde{\xi} =  2\chi_{\star}$ into (\ref{rs2}) as given in the main text, and using the self consistent equation for  $\chi_{\star}$ we obtain
\begin{equation}
    v(1-\zeta) = \mathbb{E}\Bigg[ \frac{2\cosh^2(\chi_{\star})}{2\cosh^2(\chi_{\star}) + u^2}\Bigg] 
\end{equation}
where we used that 
\begin{equation}
    \ddot{\tilde{u}}(x) = -\frac{1}{2\cosh(x/2)}
\end{equation}
in (\ref{lin:rs2}).
Hence we get that
\begin{equation}
    \zeta = \mathbb{E}\Bigg[ \frac{\tilde{\tau}}{2\cosh^2(\chi_{\star}) +\tilde{\tau}}\Bigg]. 
\end{equation}
Since we want to solve the RS equations with $\zeta$ as control parameter, we can as well re-write the equation above as 
\begin{equation}
    \tilde{\tau} = \zeta\bigg(\mathbb{E}\Bigg[ \frac{1}{2\cosh^2(\chi_{\star}) + \tilde{\tau}}\Bigg]\bigg)^{-1}
\end{equation}
Using 
the definition of $\Tilde{\xi}$, we obtain by substitution in (\ref{rs1}) that 
\begin{equation}
    v^2 \zeta = \tau^2 \mathbb{E}\Big[\Big(\Tilde{\xi} - v/\sigma Q)\Big)^2] =  \tau^2/\sigma^2 \mathbb{E}\Big[\tanh^2(\chi_{\star})\Big]
\end{equation}
hence 
\begin{equation}
\label{v^2}
    \frac{v^2}{\sigma^2} \zeta = \tilde{\tau}^2  \mathbb{E}\Big[\tanh^2(\chi_{\star})\Big] .
\end{equation}
Equations (\ref{lin:rs4}) and (\ref{lin:rs5}) can be easily obtained by substitution, using $\mathbb{E}[Z]=0$ and 
\begin{equation}
    \dot{\tilde{u}}(x) = - \tanh(x).
\end{equation}

\section{Details for the Weibull model}
\label{appendix:weibull}
Since 
\begin{equation}
    \mathbf{X}'\bbeta_0 \overset{d}{=} \theta_0 Z_0 , \quad \theta_0 := \|\bm{\Sigma}_0^{1/2}\bbeta_0\|, \ Z_0 \sim\mathcal{N}(0,1).
\end{equation}
we have that 
\begin{eqnarray}
    \xi&=& {\rm prox}_{-\tau u(.,T,\phi,\rho)}(k\theta_0Z_0+vQ) \nonumber \\
    &=&k\theta_0Z_0+vQ + \tau -  W_0\Big(\tau \rme^{\tau + \rho/\rho_0 \log Z +\phi- \rho/\rho_0\phi_0 + (k-\rho/\rho_0)\theta_0Z_0 + vQ}\Big). \nonumber 
\end{eqnarray}
Substitution of $\xi$ in (\ref{rs3}) gives 
\begin{equation}
    k\theta_0 = k\theta_0 - \mathbb{E}_{Z,\Delta,Z_0,Q} \Big[Z_0 W_0\Big(\tau \rme^{\tau + \rho/\rho_0 \log Z +\phi- \rho/\rho_0\phi_0 +  (k-\rho/\rho_0)\theta_0Z_0 +vQ} \Big) \Big]  
\end{equation}
using 
\begin{align}
    &\mathbb{E}_{Z,\Delta,Z_0,Q} \Big[Z_0 W_0\Big(\tau \rme^{\tau + \rho/\rho_0 \log Z +\phi- \rho/\rho_0\phi_0 +  (k-\rho/\rho_0)\theta_0Z_0 +vQ} \Big) \Big] \\
    &=(k -\rho/\rho_0)\theta_0\mathbb{E}_{Z,\Delta,Z_0,Q} \Bigg[\frac{W_0\Big(\tau \rme^{\tau + \rho/\rho_0 \log Z +\phi- \rho/\rho_0\phi_0 +  (k-\rho/\rho_0)\theta_0Z_0 +vQ} \Big)}{1+W_0\Big(\tau \rme^{\tau + \rho/\rho_0 \log Z +\phi- \rho/\rho_0\phi_0 +  (k-\rho/\rho_0)\theta_0Z_0 +vQ} \Big)} \Bigg]\nonumber
\end{align}
we obtain 
\begin{equation}
\label{kappa}
    k = \rho/\rho_0 
\end{equation}
which in turn simplifies greatly the following equations, because it causes  $Z_0$ to disappear. 
By substituting $\xi_{\star}$ into (\ref{rs1}) we have
\begin{equation}
    v^2 \zeta = \mathbb{E}_{Z,\Delta,Z_0,Q}\Bigg[ \bigg(\tau-  W_0\Big(\tau \rme^{\tau + \rho/\rho_0 \log Z +\phi- \rho/\rho_0\phi_0 + vQ} \Big)\bigg)^2\Bigg].
\end{equation}
Substitution of $\xi_{\star}$ into (\ref{rs2}), followed by (Gaussian) integration by parts (also known as Stein's lemma), gives
\begin{equation}
    \zeta = \mathbb{E}_{Z,\Delta,Z_0,Q}\Bigg[\frac{ W_0\Big(\tau \rme^{\tau + \rho/\rho_0 \log Z +\phi- \rho/\rho_0\phi_0 + vQ} \Big)}{1 +  W_0\Big(\tau \rme^{\tau + \rho/\rho_0 \log Z +\phi- \rho/\rho_0\phi_0 + vQ} \Big)} \Bigg]
\end{equation}
where we used  
\begin{equation}
    \frac{\partial}{\partial Q} W_0\Big(\tau \rme^{\tau + \rho/\rho_0 \log Z +\phi- \rho/\rho_0\phi_0 + vQ} \Big) = \frac{W_0\Big(\tau \rme^{\tau + \rho/\rho_0 \log Z +\phi- \rho/\rho_0\phi_0 + vQ} \Big)}{1 + W_0\Big(\tau \rme^{\tau + \rho/\rho_0 \log Z +\phi- \rho/\rho_0\phi_0 + vQ} \Big)} 
\end{equation}
which can be verified by the definition of $W_0$.

The equations for the nuisance parameters are obtained by solving the maximisation condition, i.e..\ by setting the expectation of the partial derivatives with respect to $\phi- \rho/\rho_0\phi_0$ and $\rho/\rho_0$ to zero.
Since the partial derivative of (\ref{ll_weibull}) with respect to $\phi$ is
\begin{equation}
    \frac{\partial}{\partial \phi} u (\xi_{\star},\phi) = 1 - \rme^{ \rho/\rho_0 \log Z + \phi- \rho/\rho_0\phi_0+ \xi_{\star}-\theta_0 Z_0 }
\end{equation}
taking the expectation and setting the result to zero gives 
\begin{equation}
    \tau = \mathbb{E}\Big[W_0\Big(\tau \rme^{\tau + \rho/\rho_0 \log Z +\phi- \rho/\rho_0\phi_0 + vQ} \Big)\Big]. 
\end{equation}
Doing the same for $\rho/\rho_0$ gives 
\begin{equation}
    \rho_0/\rho = \gamma_E +  \frac{1}{\tau}\mathbb{E}\Big[\log Z \ W_0\Big(\tau \rme^{\tau + \rho/\rho_0 \log Z +\phi- \rho/\rho_0\phi_0 + vQ} \Big)\Big]
\end{equation}
where the Euler-Mascheroni constant $\gamma_E$ is defined as 
\begin{equation}
    \gamma_E := \mathbb{E}\big[-\log Z\big].
\end{equation}

\section{Details for the Logit regression model}
\label{appendix:logit}
Substitution of $\xi = {\rm prox}_{-\tau u(.,T,)}(vQ+k \theta_0Z_0) = \chi_{\star} - \phi$ into (\ref{rs2}) as given in the main text leads to
\begin{equation}
    v(1-\zeta)  = \mathbb{E}\Bigg[ \frac{\cosh^2(\chi_{\star})}{\cosh^2(\chi_{\star}) + \tau}\Bigg] 
\end{equation}
where we used  
\begin{equation}
    \ddot{u}(x,T,\phi) = \frac{1}{\cosh(x + \phi)}.
\end{equation}
Hence 
\begin{equation}
    \zeta = \mathbb{E}\Bigg[ \frac{\tau}{\cosh^2(\chi_{\star}) + \tau}\Bigg]. 
\end{equation}
By the definition of $\chi$, we obtain from (\ref{rs1}) that 
\begin{equation}
    v^2 \zeta = \tau^2\mathbb{E}\Big[\Big(\dot{u}(\xi_{\star})\Big)^2\Big] = \tau^2\mathbb{E}\Big[\Big( T + \tanh(\chi_{\star})\Big)^2\Big] 
\end{equation}
where we used 
\begin{equation}
    \dot{u}(x,T,\phi) = - T - \tanh(x+\phi).
\end{equation}
Similarly we obtain that 
\begin{equation}
    \frac{\partial}{\partial \phi} u(\xi+\phi) =  - T - \tanh(\chi_{\star})
\end{equation}
so taking the expectation and setting the result to zero we get 
\begin{equation}
    \mathbb{E}\big[T\big] + \mathbb{E}\big[\tanh(\chi_{\star})\big] = 0 
\end{equation}
which in turn implies
\begin{equation}
    \phi = \mathbb{E}\big[\chi_{\star}\big].  
\end{equation}
\end{appendix}

\end{document}